\documentclass[final,1p,times]{elsarticle}
\journal{Mediterranean Journal of Mathematics}
\usepackage{hyperref}
\usepackage{tikz-cd}
\usepackage{extpfeil}
\usepackage{amssymb}
\usepackage{amsthm}
\usepackage{latexsym}
\usepackage{amsmath}
\usepackage{xcolor}
\usepackage{graphicx}
\usepackage{indentfirst}
\usepackage{mathrsfs}
\allowdisplaybreaks

\newtheorem{theorem}{\color{black}\indent \textbf{Theorem}}[section]
\newtheorem{lemma}{\color{black}\indent Lemma}[section]
\newtheorem{proposition}{\color{black}\indent Proposition}[section]
\newtheorem{definition}{\color{black}\indent Definition}[section]
\newtheorem{remark}{\color{black}\indent Remark}[section]
\newtheorem{corollary}{\color{black}\indent Corollary}[section]
\newtheorem{example}{\color{black}\indent Example}[section]

\begin{document}

\begin{frontmatter}

\title{Canonical and Canonoid Transformations of Hamiltonian Systems
on $q$-Cosymplectic and $q$-Contact Manifolds}
\author[a]{Giuseppe Marmo}
\author[b]{Cristina Sard\'on}
\author[c]{Xuefeng Zhao\corref{cor1}}

\address[a]{Dipartimento di Fisica ``Ettore Pancini'', Universit\`a di Napoli Federico II, and INFN, Sezione di Napoli, Complesso Universitario di Monte S.~Angelo, Via Cintia, 80126 Napoli, Italy}
\address[b]{Departamento de Matem\'atica Aplicada, Universidad Polit\'ecnica de Madrid, Escuela de Edificaci\'on. Av. Juan de Herrera 6, 28040 Madrid, Spain}
\address[c]{College of Mathematics, Jilin University, Changchun 130012, P. R. China}

\cortext[cor1]{Corresponding author. Email address: \texttt{zhaoxuef@jlu.edu.cn}}

\begin{abstract}
$q$-Cosymplectic and $q$-contact structures equip presymplectic
and precontact geometry with additional one-forms and a
normalized family of Reeb vector fields. We study canonical
and canonoid transformations relative to these data, with
particular attention to their origin in constrained
mechanical systems. In the $q$-cosymplectic and uniform
$q$-contact cases, we establish local normal forms,
characterize infinitesimal automorphisms, and obtain
Noether-type relations between symmetries and conserved
or dissipated quantities. For a one-parameter group of
canonoid transformations with generator $X$, we show that
$[X,X_H]$ is Hamiltonian and determine its Hamiltonian
function. In the uniform $q$-contact case, simultaneous
constant rescalings of the defining forms provide
canonoid transformations for every Hamiltonian.
The geometric constructions are illustrated by regular
relativistic mass-shell constraints and by global
Lagrangian descriptions of a charge in magnetic monopole
fields. These examples distinguish parameters along
constraint orbits from physical time and clarify the
role of gauge directions. They also recover the total
angular momentum of free relativistic particles and
the Poincar\'e angular momentum of the charge--monopole
system. For two monopoles, we construct a non-uniform
$2$-contact structure and prove a topological obstruction
to a uniform structure with the prescribed gauge
generators as Reeb fields. Irrationally related magnetic
charges additionally obstruct the intermediate quotient
that would produce a single contact manifold. This
example identifies a limitation of the uniform theory
and motivates its extension to non-uniform structures.
\end{abstract}

\begin{keyword}
$q$-cosymplectic manifold, $q$-contact manifold, canonoid transformation, symmetry, first integral, magnetic monopole, relativistic particle dynamics.
\end{keyword}
\end{frontmatter}

\section{Introduction}

Degenerate Lagrangians and Hamiltonian constraints lead
naturally to geometric descriptions in which the relevant
closed two-form has a nontrivial kernel. The foundations
of this approach were developed by Dirac and by Anderson
and Bergmann
\cite{Dirac1950,AndersonBergmann1951,Dirac1964}.
Symmetries, their generators, and their relation to
constraints were subsequently studied in detail within
generalized Hamiltonian dynamics
\cite{Mukunda1976,MarmoMukundaSamuel1983}.
These results provide the background for the present
work. Our purpose is to study the consequences of
specifying additional geometric data on the kernel and
on a complementary distribution, as occurs in
$q$-cosymplectic and $q$-contact geometry.

On a $q$-cosymplectic manifold, the closed two-form
$\Omega$ defines a presymplectic structure with
$q$-dimensional kernel $\mathcal R$. The closed
one-forms $\lambda_1,\dots,\lambda_q$ determine a
complementary distribution
$\xi=\bigcap_i\ker\lambda_i$ and a distinguished frame
$R_1,\dots,R_q$ of $\mathcal R$, normalized by
$\lambda_i(R_j)=\delta_{ij}$. Thus the structure includes
a choice of transverse directions and a normalization
of the characteristic directions. In the $q$-contact
case, each $\lambda_i$ is a precontact form, with
characteristic distribution
$\ker\lambda_i\cap\ker d\lambda_i$. The forms are related
by the requirement that all $d\lambda_i$ have the same
kernel $\mathcal R$ and are non-degenerate on the common
complement $\xi$. Uniformity imposes the further
condition $d\lambda_1=\cdots=d\lambda_q$.

These additional data affect the transformation theory.
Preserving a presymplectic form preserves its kernel,
but need not preserve a chosen complementary
distribution or the individual Reeb fields.
For a Hamiltonian constant along the kernel, the
choice of $\xi$ selects a unique horizontal solution
of the presymplectic Hamiltonian equation.
Similarly, preserving one precontact form does not
in general preserve an entire $q$-contact structure.
We therefore formulate canonical transformations,
Hamiltonian vector fields, and canonoid transformations
with explicit reference to the defining one-forms.
The physical examples below explain how such data can
arise and why their local and global properties require
separate consideration.

\subsection*{Geometric and dynamical background}
Two earlier constructions are directly relevant to the present
work. Dubrovin, Giordano, Marmo and Simoni
\cite{DubrovinGiordanoMarmoSimoni1993} construct Poisson
brackets from a presymplectic form and a flat generalized
connection. The $q$-cosymplectic bracket used here is obtained
by taking the horizontal projector determined by the closed
one-forms $\lambda_i$; see
Remark~\ref{rem:presymplectic-poisson}.
Alekseevsky, Grabowski, Marmo and Michor
\cite{AlekseevskyGrabowskiMarmoMichor1994} describe symplectic
and Poisson structures on cotangent bundles of Lie groups
and principal bundles in terms of momentum mappings,
Maurer--Cartan forms, and connections.
Their principal-bundle construction provides a geometric
framework for the Cartan forms in our monopole examples;
see Remark~\ref{rem:principal-bundle-comparison}.
These works supply the underlying bracket and bundle
constructions. Our transformation theory keeps explicit
the additional ordered one-forms and the normalized Reeb
fields, and imposes compatibility with these data.

Canonical transformations are standard tools for
comparing Hamiltonian descriptions
\cite{Bogdanovic,Carinena2}.
Canonoid transformations preserve the Hamiltonian form
of a specified dynamics while allowing its Hamiltonian
description to change; their study in symplectic
geometry includes
\cite{Currie,Negri,CarinenaRanada1988,Rastelli}.
Related constructions in cosymplectic, locally conformal
symplectic, contact, and cocontact geometry connect
transformation theory with symmetries, first integrals,
and dissipation laws
\cite{AzuajeEscobar,Azuaje2,Azuaje3,
PerezAlvarez2024,Zhao2025}.

Structures defined by several one-forms have been
studied in connection with integrability,
thermodynamical models, and Hamiltonian dynamics
\cite{deLeonBajo2025,LeokSardonZhao}.
The $q$-cosymplectic structures considered here are
those of \cite{LeokSardonZhao,Leok2}, while the
$q$-contact setting is developed in
\cite{Almeida,Finamore2024,Finamore2024b,Leok2026,ZhaoLi}.
Integrability, reduction, and Noether-type results are
already available in these settings
\cite{LeokSardonZhao,Leok2,Leok2026,ZhaoLi,
deLeon2026geometric}. A broader framework allowing
several two-forms with a common kernel is discussed in
\cite{deLeon2026geometric}. We use this framework when
the forms arising from separate magnetic contributions
must be distinguished from the form governing the
total dynamics.

A related classical construction replaces the
one-parameter group of time translations by a Lie group
$G$. This idea was already considered by J.~Schwinger
in his lectures on quantum field theory
\cite{Schwinger1970}. Hamilton--Jacobi theory in this setting was
studied by Mukunda \cite{Mukunda1978} and by Marmo,
Morandi, and Mukunda
\cite{MarmoMorandiMukunda1990}.
For a Hamiltonian action of the Abelian vector group
$\mathbb R^q$ with an equivariant momentum map,
a regular momentum level is
presymplectic, with kernel generated by the independent
infinitesimal action fields. Equipping this level with
closed one-forms whose restrictions to the kernel form
a coframe gives a $q$-cosymplectic structure. The
existence of such forms globally is an additional
requirement. This construction relates the present
geometry to the classical theory of Hamiltonian group
actions. Our transformation results concern the
Abelian setting: the Reeb fields commute and generate
a local $\mathbb R^q$-action. Non-Abelian extensions require different structure
equations and will not be dealt with in this paper.

\subsection*{Physical examples and their interpretation}

\emph{Relativistic constraint dynamics.}
The no-interaction theorem \cite{CJS1963} exhibits the
restrictions imposed by the standard canonical
world-line assumptions in a single-time relativistic
particle description. Constraint formulations provide
a way to accommodate interacting particles while
retaining relativistic covariance
\cite{Dirac1964,BMMNSSZ1984}.
We consider $N$ independent generalized mass-shell
constraints on $T^*((\mathbb R^4)^N)$ and restrict to
a regular region in which these constraints
Poisson-commute. This commutativity is a stronger
assumption than the first-class property alone.
The constraint surface has dimension $7N$, and the
restricted symplectic form has a kernel of rank $N$.

Functions $\chi_1,\dots,\chi_N$ whose differentials
are independent on this kernel determine an
$N$-cosymplectic structure wherever the corresponding
transversality matrix is invertible.
Its Reeb fields are normalized linear combinations
of the constraint fields.
Their parameters describe directions along constraint
orbits. In the free-particle example, our normalization
identifies these parameters with constant multiples
of the individual proper times. For interacting
systems, an interpretation in terms of proper time
requires a separate analysis of the chosen constraints
and clock functions.

\emph{A charge in magnetic monopole fields.}
Magnetic monopoles provide a setting in which the
topology of the configuration space affects the
existence of a global potential
\cite{Dirac1931,BMSS1980,ZSNMMB1983}.
For a point monopole, the magnetic two-form on
$Q=\mathbb R^3\setminus\{0\}$ is closed but not exact.
It therefore admits no global vector potential on $Q$
and no global magnetic Lagrangian of the usual
potential form on $TQ$.
A global Lagrangian can instead be defined on $TP$,
where $P\to Q$ is a principal $U(1)$-bundle.
Its Legendre image is a constraint submanifold of
$T^*P$, on which a normalized Cartan form is contact
and its Reeb field generates the gauge action
\cite{BMSS1980,ZSNMMB1983}.
The particle still evolves with one physical time
parameter. Gauge-invariant observables descend to
the magnetic phase space $TQ$, and the lifted
Hamiltonian dynamics projects to the Lorentz force
equation.

For two monopoles, the corresponding construction uses
a principal $U(1)\times U(1)$-bundle and produces an
eight-dimensional constraint manifold.
The normalized forms associated with the separate
monopoles define a non-uniform $2$-contact structure.
The total Cartan form, which contains both magnetic
contributions, determines the physical dynamics.
There are two gauge directions and one physical
time parameter.

Two different global issues occur in this example.
If the magnetic charges have an irrational ratio,
the one-parameter subgroup along which the total
Cartan form would descend has dense orbits in the
gauge tori. Its orbit space does not give the smooth
seven-dimensional contact manifold obtained in the
rational case \cite{ZSNMMB1983}.
The quotient by the full gauge torus remains the
magnetic phase space.
Independently of this irrationality condition, we
prove that the topology of the monopole bundles
prevents a uniform $2$-contact structure whose Reeb
fields are the prescribed gauge generators.
The example therefore distinguishes the obstruction
to an intermediate contact quotient from the
obstruction to uniformity.

\subsection*{Results and scope}

We develop the transformation theory for
$q$-cosymplectic and uniform $q$-contact structures,
keeping the additional one-forms explicit.
The local normal forms and Noether-type statements
are presented as adaptations of established
geometric methods to these specified data.
The main results concern the compatibility conditions
for transformations, their infinitesimal generators,
and the geometric restrictions exhibited by the
physical examples.

In the $q$-cosymplectic case, canonical transformations
preserve both $\Omega$ and each $\lambda_i$.
We distinguish this requirement from preservation
of the Reeb frame and the complementary distribution;
see Proposition~\ref{prop:q-cosymplectic-equivalence}.
We also characterize horizontal generators of
generalized canonical transformations by locally
Hamiltonian vector fields.
For a Hamiltonian $H$, we consider the family
$E_H^{\vec a}=X_H+\sum_i a_iR_i$, where
$\vec a\in\mathbb R^q$.
The Noether theorem relates horizontal infinitesimal
symmetries to functions conserved by every member
of this family. These functions are invariant under
each Reeb field as well as under $X_H$.
This simultaneous conservation condition does not
require the different fields $E_H^{\vec a}$ to commute.

In the uniform $q$-contact case, a symmetry preserving
$H$ and all the defining one-forms gives a tuple
$(\lambda_1(V),\dots,\lambda_q(V))$ satisfying a common
dissipation law.
Uniformity also imposes a restriction: for such
strict symmetries, the differences
$\lambda_i(V)-\lambda_j(V)$ are locally constant.
We retain the tuple formulation to express
compatibility with the whole family of one-forms.

For both structures, let $X$ generate a one-parameter
group of canonoid transformations with smoothly
varying Hamiltonians $K_s$, normalized by $K_0=H$.
Writing $\kappa=\left.\frac{d}{ds}\right|_{s=0}K_s$,
we prove the necessary infinitesimal relation
$[X,X_H]=X_{X(H)-\kappa}$.
The equations determining $\kappa$ depend on the
geometric structure and, in the uniform $q$-contact
case, must hold simultaneously for all defining
one-forms.
We also give an explicit class of canonoid flows
valid for every Hamiltonian: those generated by
vector fields satisfying
$\mathcal L_X\lambda_i=\lambda_i$ for all $i$.

The examples recover familiar conservation laws
within these descriptions: the total angular momentum
of free relativistic particles and the Poincar\'e
angular momentum of a charge in a monopole field.
The two-monopole example has a different role.
Its non-uniform structure and the obstruction just
described identify a physical situation beyond the
uniform transformation theory developed here.

\subsection*{Organization of the paper}

Section~\ref{sec:prelim} recalls the geometric structures,
establishes local normal forms, and describes the
Jacobi structure associated with a uniform
$q$-contact manifold.
Section~\ref{sec:examples} develops the constraint and
monopole constructions. It also gives a local passage
from $q$-cosymplectic to uniform $q$-contact structures
under a normalization of a primitive of $\Omega$,
together with a global realization for free
relativistic particles.
Sections~\ref{sec:cosymp} and \ref{sec:contact} contain
the canonical and canonoid transformation results
and their relations to symmetries and conservation
or dissipation laws.
Section~\ref{sec:conclusions} discusses the extension
to non-uniform structures, the integration of
infinitesimal conditions, and further questions
suggested by the examples.

\section{Preliminaries}\label{sec:prelim}

All manifolds, differential forms, and vector fields are
smooth and real. We use the contraction convention
$(\iota_X\Omega)(Y)=\Omega(X,Y)$.
The purpose of this section is to fix the geometric
data and sign conventions used below, and to explain
their relation to presymplectic and precontact geometry.
We include proofs of the local normal forms and basic
Hamiltonian identities for completeness.

\subsection{A common geometric setting}

\begin{definition}[$q$-structured manifolds;
cf.~\cite{deLeon2026geometric}]
Let $n,q,m$ be positive integers and let $M$ have
dimension $2n+q$.
A \emph{$q$-structure} on $M$ consists of one-forms
$\vec\lambda=(\lambda_1,\dots,\lambda_q)$,
two-forms $\vec\omega=(\Omega_1,\dots,\Omega_m)$,
and a splitting $TM=\mathcal R\oplus\xi$ such that:
\begin{itemize}
\item[(i)] the $\lambda_i$ are pointwise linearly
independent and $\xi=\bigcap_{i=1}^q\ker\lambda_i$;
\item[(ii)] each restriction $\Omega_j|_\xi$ is
non-degenerate;
\item[(iii)] $\ker\Omega_j=\mathcal R$ for every
$j=1,\dots,m$.
\end{itemize}
\end{definition}

A manifold endowed with these data is called a
\emph{$q$-structured manifold}.
We use this terminology to collect the algebraic data
shared by the structures considered below.
No closedness condition is imposed at this stage.
The case $m=q$ is the framework of
\cite{deLeon2026geometric}; here the number of two-forms
is allowed to differ from the number of one-forms.
This is useful in Proposition~\ref{prop:nonuniform},
where $d\lambda_1$, $d\lambda_2$, and the physically
relevant form $d\theta_L$ occur together, so that
$q=2$ and $m=3$.
We write
$(M,\vec\lambda\oplus\vec\omega,\mathcal R\oplus\xi)$
for these data.
The bundles $\mathcal R$ and $\xi$ have ranks $q$
and $2n$, respectively, and are called the
\emph{Reeb distribution} and the
\emph{horizontal distribution}.
The nowhere-vanishing $q$-form
$\lambda_1\wedge\cdots\wedge\lambda_q$
is the \emph{characteristic form}.

\begin{remark}
Every $\Omega_j$ has rank $2n$.
Its restriction to $\xi$ makes $\xi$ a symplectic
vector bundle, a statement of fibrewise linear
algebra that does not require $d\Omega_j=0$.
Non-degeneracy of the individual restrictions does
not imply non-vanishing of their mixed wedge products.
For example, on $\mathbb R^4$, let
$a=dx_1\wedge dy_1$ and $b=dx_2\wedge dy_2$.
Both $a+b$ and $a-b$ are non-degenerate, whereas
$(a+b)\wedge(a-b)=0$.
\end{remark}

\begin{proposition}\label{prop:common-flat}
For each $j=1,\dots,m$, the vector bundle map
\[
\flat^j:TM\longrightarrow T^*M,\qquad
\flat^j(X)=\iota_X\Omega_j+
\sum_{k=1}^q\lambda_k(X)\lambda_k
\]
is an isomorphism.
It therefore induces an isomorphism of
$C^\infty(M)$-modules between vector fields and
one-forms.
\end{proposition}

\begin{proof}
Fix $x\in M$. If $\flat^j_x(v)=0$, evaluation on $v$
gives
$0=\sum_{k=1}^q\lambda_k(v)^2$.
Thus $v\in\xi_x$ and $\iota_v\Omega_j=0$.
Non-degeneracy of $\Omega_j|_{\xi_x}$ implies $v=0$.
The domain and codomain have the same dimension,
so $\flat^j_x$ is an isomorphism.
These maps depend smoothly on $x$, and their inverses
are smooth.
\end{proof}

\begin{corollary}\label{cor:common-reeb}
There are unique vector fields $R_1,\dots,R_q$
satisfying
\[
\iota_{R_k}\Omega_j=0,\qquad
\lambda_l(R_k)=\delta_{lk},
\qquad
j=1,\dots,m,\quad k,l=1,\dots,q.
\]
They form a global frame of $\mathcal R$ and are
called the \emph{Reeb vector fields}.
Moreover, $(\flat^j)^{-1}(\lambda_k)=R_k$ for every $j$.
\end{corollary}

\begin{proof}
At each point, consider the map
\[
\mathcal R_x\longrightarrow\mathbb R^q,\qquad
v\longmapsto
\bigl(\lambda_1(v),\dots,\lambda_q(v)\bigr).
\]
Its kernel is $\mathcal R_x\cap\xi_x=\{0\}$.
Since both spaces have dimension $q$, this map is
an isomorphism.
The inverse images of the standard basis of
$\mathbb R^q$ define the required smooth fields.
They belong to every $\ker\Omega_j$ and satisfy the
normalization conditions.
These conditions also prove uniqueness.
Finally, $\flat^j(R_k)=\lambda_k$.
\end{proof}

In particular, the chosen forms determine the
projections associated with the splitting:
\[
\operatorname{pr}_{\mathcal R}X
=\sum_{i=1}^q\lambda_i(X)R_i,\qquad
\operatorname{pr}_{\xi}X
=X-\sum_{i=1}^q\lambda_i(X)R_i.
\]
At this level of generality, the Reeb fields need
not commute. Their commutativity below follows
from the additional differential conditions.

\begin{remark}
When all the $\Omega_j$ coincide with a form $\Omega$,
the data define an almost cosymplectic structure of
order $q$ in the terminology of
\cite{deLeonBajo2025}.
If $d\Omega=0$, the structure is partially
cosymplectic of order $q$.
If the $\lambda_i$ are also closed, we use the term
$q$-cosymplectic, following
\cite{LeokSardonZhao}.
When $m=q$ and $\Omega_i=d\lambda_i$, the data define
a $q$-contact structure
\cite{Almeida,Finamore2024}.
These cases have different integrability properties
and will be treated separately.
\end{remark}

\subsection{$q$-cosymplectic manifolds}

\begin{definition}[$q$-cosymplectic manifolds
\cite{LeokSardonZhao}]\label{q-co}
A \emph{$q$-cosymplectic structure} on a manifold
$M^{2n+q}$ consists of a closed two-form $\Omega$
of rank $2n$ and closed one-forms
$\vec\lambda=(\lambda_1,\dots,\lambda_q)$ such that
\[
\Omega^n\wedge\lambda_1\wedge\cdots\wedge\lambda_q
\neq0
\]
at every point.
Equivalently, with
$\mathcal R=\ker\Omega$ and
$\xi=\bigcap_i\ker\lambda_i$, one has
$TM=\mathcal R\oplus\xi$ and
$\Omega|_\xi$ is non-degenerate.
We denote this structure by $(M,\Omega,\vec\lambda)$.
\end{definition}

The rank assumption in Definition~\ref{q-co}
is part of the definition.
Together with the volume condition, it says that
the restrictions of the $\lambda_i$ to
$\ker\Omega$ form a coframe.
Thus $(M,\Omega)$ is presymplectic, and the
one-forms specify both a complementary distribution
and a normalized frame of the kernel.
The splitting and the Reeb fields follow from
these data by Corollary~\ref{cor:common-reeb}.

\begin{proposition}\label{prop:commute}
On a $q$-cosymplectic manifold, the Reeb fields
satisfy
\[
[R_i,R_j]=0,\qquad
\mathcal L_{R_i}\Omega=0,\qquad
\mathcal L_{R_i}\lambda_j=0
\]
for all $i,j$.
Moreover, $\xi$ is integrable and $\Omega$ restricts
to a symplectic form on each leaf of $\xi$.
\end{proposition}

\begin{proof}
Cartan's formula gives
$\mathcal L_{R_i}\Omega=0$ and
$\mathcal L_{R_i}\lambda_j
=d(\lambda_j(R_i))=0$.
Consequently,
\[
\iota_{[R_i,R_j]}\Omega
=\mathcal L_{R_i}(\iota_{R_j}\Omega)
-\iota_{R_j}(\mathcal L_{R_i}\Omega)=0,
\]
while
$\lambda_k([R_i,R_j])
=R_i(\lambda_k(R_j))
-(\mathcal L_{R_i}\lambda_k)(R_j)=0$.
Thus $[R_i,R_j]$ belongs to
$\mathcal R\cap\xi=\{0\}$.

If $U,V$ take values in $\xi$, then
$\lambda_i([U,V])=-d\lambda_i(U,V)=0$.
Hence $\xi$ is integrable.
The restriction of $\Omega$ to a leaf is closed
and non-degenerate, so it is symplectic.
\end{proof}

The Reeb fields generate a local
$\mathbb R^q$-action.
A global action requires their completeness.
In particular, the closedness assumptions in
Definition~\ref{q-co} force the Reeb algebra to
be Abelian.

\begin{theorem}[Darboux theorem]\label{Darb}
Let $(M,\Omega,\vec\lambda)$ be a $q$-cosymplectic
manifold of dimension $2n+q$.
Around every point there are coordinates
$(q^1,\dots,q^n,p_1,\dots,p_n,t_1,\dots,t_q)$ such that
\[
\Omega=\sum_{i=1}^n dq^i\wedge dp_i,\qquad
\lambda_j=dt_j,\qquad
R_j=\frac{\partial}{\partial t_j}.
\]
\end{theorem}

\begin{proof}
Fix $x_0\in M$.
After restricting to a neighbourhood of $x_0$,
write $\lambda_j=dt_j$ with $t_j(x_0)=0$.
The differentials $dt_j$ are independent, so
$\Sigma_0=\{t_1=\cdots=t_q=0\}$ is locally a
$2n$-dimensional submanifold with
$T\Sigma_0=\xi|_{\Sigma_0}$.
If $\iota:\Sigma_0\hookrightarrow M$ is the inclusion,
then $\iota^*\Omega$ is symplectic.
Choose symplectic Darboux coordinates $(q,p)$
on a neighbourhood in $\Sigma_0$.

Let $\phi^j_s$ denote the local flow of $R_j$.
For $x\in\Sigma_0$ and sufficiently small
$s=(s_1,\dots,s_q)$, define
\[
\Phi(x,s)
=\phi^1_{s_1}\circ\cdots\circ\phi^q_{s_q}(x).
\]
At $(x_0,0)$ its differential is
$(v,a)\mapsto v+\sum_j a_jR_j(x_0)$, which is an
isomorphism because
$T_{x_0}M=T_{x_0}\Sigma_0\oplus\mathcal R_{x_0}$.
Thus, after shrinking the domain, $\Phi$ is a
diffeomorphism onto a neighbourhood of $x_0$.

By Proposition~\ref{prop:commute}, the flows commute
and $\Phi_*\partial_{s_j}=R_j$.
Since $R_j(t_k)=\delta_{jk}$ and the $t_k$ vanish
on $\Sigma_0$, one has $t_k\circ\Phi=s_k$.
Extend the functions $(q,p)$ from $\Sigma_0$ by
making them constant along the Reeb flows.
Then $(q,p,t)$ are coordinates with
$R_j=\partial_{t_j}$ and $\lambda_j=dt_j$.

Finally, $\iota_{\partial_{t_j}}\Omega=0$ excludes
all $dt_j$-components of $\Omega$, and
$\mathcal L_{\partial_{t_j}}\Omega=0$ makes the
remaining coefficients independent of $t$.
Its expression is therefore the pull-back of its
Darboux expression on $\Sigma_0$.
\end{proof}

\begin{definition}
For $H\in C^\infty(M)$, its \emph{Hamiltonian vector
field} is the unique field $X_H$ satisfying
\[
\iota_{X_H}\Omega
=dH-\sum_{i=1}^qR_i(H)\lambda_i,\qquad
\lambda_i(X_H)=0,\quad i=1,\dots,q.
\]
The associated bracket is
$\{f,g\}:=\Omega(X_f,X_g)$.
\end{definition}

The right-hand side of the contraction equation
annihilates $\mathcal R$.
Existence and uniqueness of $X_H\in\xi$ therefore
follow from non-degeneracy of $\Omega|_\xi$.
In the coordinates of Theorem~\ref{Darb},
\[
X_f
=\sum_{i=1}^n
\left(
\frac{\partial f}{\partial p_i}\frac{\partial}{\partial q^i}
-\frac{\partial f}{\partial q^i}\frac{\partial}{\partial p_i}
\right),
\qquad
\{f,g\}
=\sum_{i=1}^n
\left(
\frac{\partial f}{\partial q^i}\frac{\partial g}{\partial p_i}
-\frac{\partial f}{\partial p_i}\frac{\partial g}{\partial q^i}
\right).
\]
In particular, $f\mapsto X_f$ is $\mathbb R$-linear.

\begin{proposition}\label{prop:qcos-poisson}
The bracket $\{\cdot,\cdot\}$ is Poisson, with
symplectic leaves given by the leaves of $\xi$.
For all $f,g\in C^\infty(M)$,
\[
\{f,g\}=X_g(f),\qquad
[X_f,X_g]=-X_{\{f,g\}},\qquad
[R_i,X_f]=X_{R_i(f)}.
\]
\end{proposition}

\begin{proof}
By the defining equations and the horizontality of
$X_g$,
\[
\Omega(X_f,X_g)
=df(X_g)-\sum_iR_i(f)\lambda_i(X_g)
=X_g(f).
\]
In Darboux coordinates, this is the canonical
Poisson bracket in $(q,p)$, with $t$ as parameters.
It satisfies the Leibniz and Jacobi identities,
and its characteristic distribution is $\xi$.

For any smooth function $h$, the Jacobi identity gives
\[
[X_f,X_g](h)
=\{\{h,g\},f\}-\{\{h,f\},g\}
=\{h,\{g,f\}\}
=-X_{\{f,g\}}(h).
\]
The last identity follows by differentiating the
coordinate expression of $X_f$ with respect to
$t_i$, since $R_i=\partial_{t_i}$.
\end{proof}

\begin{remark}[Relation to presymplectic Poisson brackets]
\label{rem:presymplectic-poisson}
The bracket above is a special case of the construction in
\cite[Sec.~4]{DubrovinGiordanoMarmoSimoni1993}.
Indeed, the additional one-forms determine the projector
\[
Q=\operatorname{id}_{TM}
  -\sum_{i=1}^q R_i\otimes\lambda_i,
\]
with $Q^2=Q$, $\ker Q=\ker\Omega=\mathcal R$, and
$\operatorname{Im}Q=\xi$.
This is the generalized connection denoted by $A$ in
that reference.

Their Eqs.~(4.4)--(4.5) associate to a one-form $\alpha$
the unique vector field satisfying
$\iota_{X_\alpha}\Omega=Q^*\alpha$ and
$QX_\alpha=X_\alpha$, where $Q^*\alpha=\alpha\circ Q$.
Taking $\alpha=df$ gives
\[
\iota_{X_f}\Omega
=df-\sum_{i=1}^q R_i(f)\lambda_i,\qquad
X_f\in\Gamma(\xi).
\]
Thus, with our bracket convention, their construction
gives exactly $\{f,g\}=\Omega(X_f,X_g)$.

In 
\cite{DubrovinGiordanoMarmoSimoni1993}, the authors also state that this
bracket satisfies the Jacobi identity if and only if the
generalized connection is flat.
The vertical distribution is involutive because
$d\Omega=0$, while the horizontal curvature vanishes:
\[
(\operatorname{id}_{TM}-Q)[QU,QV]
=-\sum_{i=1}^q d\lambda_i(QU,QV)R_i=0.
\]
Hence the closedness of the $\lambda_i$ supplies precisely
the required flatness.

On the algebra
$\mathcal A_\Omega
=\{f\in C^\infty(M):df|_{\mathcal R}=0\}$,
the bracket agrees with the intrinsic presymplectic
bracket.
For such $f$, the solutions of
$\iota_{Y_f}\Omega=df$ are
$Y_f=X_f+\sum_i a_iR_i$, and their pairings under
$\Omega$ are independent of the functions $a_i$.
The extension to all smooth functions depends on the
chosen complement $\xi$.
No global smooth quotient by $\mathcal R$ is required.

The full $q$-cosymplectic structure also specifies the
ordered closed one-forms and their dual Reeb frame.
Preservation of these data together with $\Omega$ is
stronger than preservation of the induced Poisson bracket.
In particular, Cartan's formula gives
\[
\mathcal L_{X_f}\lambda_i=0,\qquad
\mathcal L_{X_f}\Omega
=-\sum_{i=1}^q d(R_i(f))\wedge\lambda_i.
\]
These identities will be used in the transformation theory.
\end{remark}

\subsection{$q$-contact manifolds}

\begin{definition}[$q$-contact manifolds
\cite{Almeida,Finamore2024,Finamore2024b,ZhaoLi}]
A \emph{$q$-contact structure} on $M^{2n+q}$ consists
of pointwise linearly independent one-forms
$\vec\lambda=(\lambda_1,\dots,\lambda_q)$ and
a splitting $TM=\mathcal R\oplus\xi$ such that
$\xi=\bigcap_i\ker\lambda_i$,
$\ker d\lambda_i=\mathcal R$, and
$d\lambda_i|_\xi$ is non-degenerate for every $i$.
It is \emph{uniform} if
$d\lambda_1=\cdots=d\lambda_q=:\Omega$.
\end{definition}

These data fit the common setting above with
$m=q$ and $\Omega_i=d\lambda_i$.
Thus there are unique Reeb fields $R_1,\dots,R_q$
such that $\lambda_i(R_j)=\delta_{ij}$ and
$\iota_{R_j}d\lambda_i=0$ for every $i,j$.
The next statement does not require uniformity.

\begin{proposition}[{\cite{Almeida,Finamore2024}}]\label{P1}
On every $q$-contact manifold, the Reeb fields
form a commuting frame of $\mathcal R$ and satisfy
\[
\mathcal L_{R_i}\lambda_j=0,\qquad
\mathcal L_{R_i}d\lambda_j=0.
\]
In the uniform case, in particular,
$\mathcal L_{R_i}\Omega=0$.
\end{proposition}

\begin{proof}
Cartan's formula gives
$\mathcal L_{R_i}\lambda_j
=d(\delta_{ji})+\iota_{R_i}d\lambda_j=0$,
and exterior differentiation gives the second
identity.
For every $k$,
$\lambda_k([R_i,R_j])=0$.
Also,
\[
\iota_{[R_i,R_j]}d\lambda_k
=\mathcal L_{R_i}(\iota_{R_j}d\lambda_k)
-\iota_{R_j}(\mathcal L_{R_i}d\lambda_k)=0.
\]
Hence $[R_i,R_j]$ belongs to both $\xi$ and
$\mathcal R$, and therefore vanishes.
\end{proof}

\begin{remark}[Relation to precontact geometry]
Each $\lambda_i$ is a precontact form of constant
class $2n+1$.
Its characteristic distribution is
\[
\mathcal C_i
:=\ker\lambda_i\cap\ker d\lambda_i
=\operatorname{span}\{R_j:j\neq i\},
\]
of rank $q-1$.
Thus $\mathcal C_i$ must be distinguished from the
$q$-dimensional common kernel $\mathcal R$.
The full $q$-contact structure specifies a family
of precontact forms together with their common
kernel, their common complement, and its normalized
Reeb frame.
When $q=1$, the form is contact and
$\mathcal C_1=\{0\}$.
\end{remark}

We now assume uniformity.
The following normal form is recalled from
\cite{Finamore2024}; we include a proof adapted
to our conventions.

\begin{proposition}[Darboux theorem]\label{prop:darboux-contact}
Let $(M,\vec\lambda)$ be a uniform $q$-contact
manifold of dimension $2n+q$.
Around every point there are coordinates
$(q^1,\dots,q^n,p_1,\dots,p_n,z_1,\dots,z_q)$ such that
\[
\lambda_j=dz_j-\sum_{i=1}^n p_i\,dq^i,\qquad
\Omega=\sum_{i=1}^n dq^i\wedge dp_i,\qquad
R_j=\frac{\partial}{\partial z_j}.
\]
\end{proposition}

\begin{proof}
By Proposition~\ref{P1}, $\mathcal R$ is integrable.
On a sufficiently small foliation chart $U$, let
$\pi:U\to B$ be a submersion onto a
$2n$-dimensional manifold with
$\ker T\pi=\mathcal R|_U$.
The form $\Omega$ annihilates $\mathcal R$ and is
invariant under the Reeb fields.
It therefore descends to a form $\omega$ on $B$,
with $\Omega=\pi^*\omega$.

The form $\omega$ is closed.
If a vector in $TB$ is in its kernel, any lift of
that vector lies in $\ker\Omega=\ker T\pi$.
Thus $\omega$ is non-degenerate.
Choose Darboux coordinates $(q,p)$ on $B$ and set
$\theta=-\sum_i p_i\,dq^i$, so that
$d\theta=\omega$.
Since $d(\lambda_j-\pi^*\theta)=0$, after shrinking
$U$ there are functions $z_j$ such that
$\lambda_j-\pi^*\theta=dz_j$.

The differentials of the pulled-back functions
$(q,p)$ annihilate $\mathcal R$, whereas
$dz_j(R_k)=\delta_{jk}$.
Consequently, the differentials of $(q,p,z)$ are
linearly independent and these functions form a
coordinate system.
The stated expressions for $\lambda_j$, $\Omega$,
and $R_j$ follow.
Only a local quotient has been used in this argument.
\end{proof}

\begin{proposition}[Contact structures on leaves]
\label{prop:uniform-contact-leaves}
Set $E=\sum_{i=1}^qR_i$.
For $j=2,\dots,q$, the forms
$\beta_j=\lambda_j-\lambda_1$ are closed and
pointwise linearly independent.
Their common kernel is the integrable distribution
\[
\mathcal D
:=\bigcap_{j=2}^q\ker\beta_j
=\xi\oplus\operatorname{span}\{E\}.
\]
Every leaf $L$ of $\mathcal D$ carries the contact
form
$\alpha_L=\lambda_1|_L=\cdots=\lambda_q|_L$,
whose Reeb field is $E|_L$.
For $q=1$, the intersection is understood as $TM$.
\end{proposition}

\begin{proof}
Uniformity gives $d\beta_j=0$.
If $\sum_{j=2}^q c_j\beta_j=0$, evaluation on
$R_k$, for $k\geq2$, gives $c_k=0$.
Thus the $\beta_j$ are independent.
A vector belongs to their common kernel exactly
when all its values under the $\lambda_i$ coincide.
Using the splitting
$TM=\xi\oplus\mathcal R$, this is equivalent to
being of the form $Y+aE$ with $Y\in\xi$.
Since the $\beta_j$ are closed, $\mathcal D$ is
integrable.

On a leaf $L$, the forms $\lambda_i|_L$ coincide.
Moreover, $\alpha_L(E)=1$,
$\iota_Ed\alpha_L=0$, and
$d\alpha_L|_\xi=\Omega|_\xi$ is non-degenerate.
Hence $\alpha_L\wedge(d\alpha_L)^n$ is nowhere zero,
and $E|_L$ is its Reeb field.
\end{proof}

In the coordinates of
Proposition~\ref{prop:darboux-contact}, put
$w=z_1$ and $u_j=z_j-z_1$ for $j=2,\dots,q$.
Then
\[
\lambda_1=dw-\sum_i p_i\,dq^i,\qquad
\lambda_j=\lambda_1+du_j,\qquad
E=\partial_w.
\]
The leaves of $\mathcal D$ are locally the sets
$u=\mathrm{constant}$.
This gives a local description by ordinary contact
manifolds together with transverse closed forms.
The local functions $u_j$ need not exist globally,
and the description does not imply a global
product decomposition.

\begin{definition}
Let $(M,\vec\lambda)$ be uniform and let
$H\in C^\infty(M)$.
The \emph{Hamiltonian vector field} $X_H$ is the
unique field satisfying
\[
\iota_{X_H}\Omega
=dH-\sum_{i=1}^qR_i(H)\lambda_i,\qquad
\lambda_i(X_H)=-H,\quad i=1,\dots,q.
\]
The associated bracket is
\[
\{f,g\}
:=\Omega(X_f,X_g)+fE(g)-gE(f),
\qquad E=\sum_{i=1}^qR_i.
\]
The Hamiltonian system is called \emph{good} if
$R_i(H)=0$ for every $i$.
\end{definition}

To see existence and uniqueness, let $Y_H\in\xi$
be the unique solution of
$\iota_{Y_H}\Omega
=dH-\sum_iR_i(H)\lambda_i$.
Then $X_H=Y_H-HE$ satisfies both equations.
Any two solutions differ by a vector in
$\mathcal R$ annihilated by all the $\lambda_i$,
and hence coincide.
In particular, $X_H$ is tangent to $\mathcal D$.
On each leaf $L$, its restriction is the usual
contact Hamiltonian field of $H|_L$, with convention
$\alpha_L(X_H)=-H|_L$.

In Darboux coordinates, the expression is
\[
\begin{aligned}
X_H={}&
\sum_{i=1}^n
\left[
\frac{\partial H}{\partial p_i}\partial_{q^i}
-\left(
\frac{\partial H}{\partial q^i}
+p_i\sum_{j=1}^q\frac{\partial H}{\partial z_j}
\right)\partial_{p_i}
\right]\\
&+\left(
\sum_{i=1}^n p_i\frac{\partial H}{\partial p_i}-H
\right)\sum_{j=1}^q\partial_{z_j}.
\end{aligned}
\]
The map $H\mapsto X_H$ is $\mathbb R$-linear and
injective, since $\lambda_i(X_H)=-H$.
Cartan's formula and the contraction equation give
\[
\mathcal L_{X_H}\lambda_i
=-\sum_{j=1}^qR_j(H)\lambda_j,\qquad
X_H(H)=-H\,E(H).
\]
Thus a good Hamiltonian is conserved along its
flow, and its Hamiltonian field preserves every
$\lambda_i$.
For all $H$, one also has $X_H(u_j)=0$ in the
local coordinates above.

\begin{remark}[Jacobi structure and signs]\label{rem:jacobi}
The defining equations give
$\Omega(X_f,X_g)=X_g(f)+gE(f)$, and consequently
\[
\{f,g\}=X_g(f)+fE(g)
=-X_f(g)-gE(f).
\]
The bilinear expression $\Omega(X_f,X_g)$ depends
pointwise only on $df$ and $dg$, since the Reeb
components of $X_f$ and $X_g$ are annihilated by
$\Omega$.
It therefore defines a smooth bivector $\Lambda$
by $\Lambda(df,dg)=\Omega(X_f,X_g)$.
In Darboux coordinates,
\[
\Lambda
=\sum_{i=1}^n
\bigl(\partial_{q^i}+p_iE\bigr)\wedge\partial_{p_i},
\qquad
\{f,g\}=\Lambda(df,dg)+fE(g)-gE(f).
\]

We verify the Lie property using
Proposition~\ref{prop:uniform-contact-leaves}.
On a leaf $L$, write
$h=f|_L$, $k=g|_L$, and $\alpha=\alpha_L$.
The restricted Hamiltonian fields satisfy
$\mathcal L_{X_h}\alpha=-E(h)\alpha$, so their
commutator is again a contact vector field.
Furthermore,
\[
\begin{aligned}
\alpha([X_h,X_k])
&=X_h(\alpha(X_k))
-(\mathcal L_{X_h}\alpha)(X_k)\\
&=-X_h(k)-kE(h)
=\{h,k\}.
\end{aligned}
\]
A contact vector field is uniquely determined by
its contraction with $\alpha$.
With our sign convention, this proves
$[X_h,X_k]=-X_{\{h,k\}}$ on each leaf.
Since the fields are tangent to $\mathcal D$,
the identity $[X_f,X_g]=-X_{\{f,g\}}$ holds on $M$.
The Jacobi identity for vector fields now gives
$X_J=0$, where
$J=\{\{f,g\},h\}+\{\{g,h\},f\}+\{\{h,f\},g\}$.
Injectivity of the Hamiltonian assignment implies
$J=0$.
Thus $(\Lambda,E)$ is a Jacobi structure.

The bracket is not Poisson: $\{1,f\}=E(f)$, and
$E$ is nowhere zero.
For example, when $q=1$ and
$\lambda=dz-p\,dq$, our conventions give
$\{q,p\}=1$, $\{p,z\}=0$, $\{z,q\}=-q$, and
$\{1,z\}=1$.
These identities fix the relative signs of the
Hamiltonian equations and the bracket.
\end{remark}

\begin{remark}[The additional one-form data]
The Jacobi structure describes the contact
Hamiltonian dynamics on the leaves of $\mathcal D$,
but it does not determine the entire ordered family
$(\lambda_1,\dots,\lambda_q)$.
For example, take $q=2$ in the coordinates
$(q,p,w,u)$ above and write
$\alpha=dw-\sum_i p_i\,dq^i$.
Then $\lambda_1=\alpha$ and $\lambda_2=\alpha+du$.
The local diffeomorphism
$F(q,p,w,u)=(q,p,w,2u)$ preserves $\Lambda$ and $E$,
whose expressions involve only $(q,p,w)$.
However,
$F^*\lambda_2=\alpha+2du\neq\lambda_2$.
Thus preserving the Jacobi structure is weaker than
preserving the defining one-forms individually.
This distinction is part of the transformation
problem studied in Section~\ref{sec:contact}.
\end{remark}

\section{The motivating examples}\label{sec:examples}

We now explain how the geometric data of
Section~\ref{sec:prelim} arise in constrained mechanics.
We first describe the passage from a regular constraint
surface to a $q$-cosymplectic structure, including an
explicit construction for the Abelian vector group.
We then consider relativistic particle constraints and
global Lagrangian descriptions of magnetic monopoles.
The underlying presymplectic constructions belong to
the classical theory of constrained systems
\cite{MarmoMukundaSamuel1983,BMMNSSZ1984,ZSNMMB1983}.
Our purpose is to identify the additional one-forms,
their normalization, and the resulting restrictions on
the transformation theory.

The interpretation of the characteristic directions
depends on the system. In relativistic constraint
dynamics they parametrize constraint orbits; in the
free-particle example these parameters are proportional
to the individual proper times. For a charge in a
magnetic field, the physical motion has a single time
parameter, while the Reeb fields of the lifted
description generate internal gauge transformations.
In particular, the integer $q$ does not by itself
count physical times or gauge directions that cannot
be quotiented out.

On a symplectic manifold we use
$\iota_{X_f}\omega=df$ and
$\{f,g\}_\omega=\omega(X_f,X_g)=X_g(f)$, consistently
with Section~\ref{sec:prelim}.
When $\theta$ is a Liouville or Cartan form, we retain
the convention $\omega=d\theta$ used in the mechanical
examples below. The variational equations then have
the form $\iota_\Gamma d\theta=-dE$.

\subsection{Constraint surfaces and Abelian group actions}
\label{subsec:constraint-origin}

The following construction separates the presymplectic
structure determined by the constraints from the
additional choice of clock functions.

\begin{proposition}\label{prop:constraints-qcos}
Let $(P,\omega)$ be symplectic of dimension $2n+2q$.
Let $C:P\to\mathbb R^q$ have components
$C_1,\dots,C_q$.
Suppose that $c$ is a regular value, that
$M=C^{-1}(c)$ is nonempty, and that
$\{C_a,C_b\}_\omega|_M=0$ for all $a,b$.
Write $\iota:M\hookrightarrow P$ and
$\Omega=\iota^*\omega$.
Then $\Omega$ has rank $2n$, and
\[
\mathcal R:=\ker\Omega
 =\operatorname{span}\{X_{C_1}|_M,\dots,X_{C_q}|_M\}.
\]
Let $\chi_1,\dots,\chi_q$ be functions on an open
subset $U\subset M$ such that the matrix
$A_{ab}=d\chi_a(X_{C_b}|_M)$ is invertible. The functions $\chi_a$ thus provide local gauge-fixing conditions
for the first-class constraints $C_a-c_a=0$. More precisely, if
$\widetilde\chi_a$ are local extensions of $\chi_a$ to $P$, then
the constraints $C_a-c_a=0$ and $\widetilde\chi_a-\tau_a=0$,
$a=1,\dots,q$, form a second-class system along each nonempty
slice $\Sigma_\tau=\{x\in U:\chi(x)=\tau\}$,
where $\chi=(\chi_1,\dots,\chi_q)$. 
Then $(U,\Omega|_U,d\chi_1,\dots,d\chi_q)$ is
$q$-cosymplectic, with Reeb fields
\[
R_a=\sum_{b=1}^q(A^{-1})_{ba}\,X_{C_b}|_U.
\]
Such functions exist locally around every point of $M$.
\end{proposition}

\begin{proof}
Regularity gives $\dim M=2n+q$ and linear
independence of the constraint fields along $M$.
At each $x\in M$, the symplectic orthogonal of $T_xM$
is spanned by $X_{C_a}(x)$.
The assumption on the brackets makes these fields
tangent to $M$. Thus $M$ is coisotropic, and
$\ker\Omega_x=(T_xM)^\omega$ has dimension $q$.

Invertibility of $A$ means that the restrictions of
$d\chi_1,\dots,d\chi_q$ to $\mathcal R$ form a coframe. Since $X_{C_b}$ is tangent to $M$, we have
$\{\widetilde\chi_a,C_b\}_\omega|_U=A_{ab}$ for any such
extensions. Ordering the constraints as
$(C-c,\widetilde\chi-\tau)$, their Poisson bracket matrix
along $\Sigma_\tau$ is
\[
\begin{pmatrix}
0 & -A^{\mathsf T}\\
A & B
\end{pmatrix},
\qquad
B_{ab}=\{\widetilde\chi_a,\widetilde\chi_b\}_\omega
       \big|_{\Sigma_\tau}.
\]
Its determinant is $(\det A)^2$, so the combined constraints
are second class. 
Consequently,
$TU=\mathcal R|_U\oplus\bigcap_a\ker d\chi_a$.
The restriction of $\Omega$ to this complement is
non-degenerate, and all the defining forms are closed.
The displayed formula for $R_a$ follows from
$d\chi_b(R_a)=\delta_{ba}$.

Finally, at any point choose $q$ covectors whose
restrictions to $\mathcal R$ are independent, and
realize them as differentials of local functions.
The corresponding matrix remains invertible after
shrinking the neighbourhood.
\end{proof}

The hypothesis on the brackets is the first-class
condition for these regular constraints on $M$.
It does not assert strong Poisson commutativity in
a neighbourhood of $M$.
For example, brackets of the form
$\{C_a,C_b\}_\omega
=\sum_d f_{ab}^{d}(C_d-c_d)$ satisfy the hypothesis,
although the restricted constraint fields need not
commute.
The normalized Reeb fields in
Proposition~\ref{prop:constraints-qcos} do commute,
by Proposition~\ref{prop:commute}.
Thus the choice of clock functions can change the
frame of the characteristic distribution; it does
not preserve a prescribed non-Abelian frame.

In particular, let $\mathbb R^q$ act in a Hamiltonian
fashion with an equivariant momentum map
$J:P\to(\mathbb R^q)^*$.
Its components Poisson-commute, and a regular momentum
level has the presymplectic structure described above.
Closed one-forms restricting to a coframe on its kernel
complete it to a $q$-cosymplectic structure.
This completion always exists locally; its global
existence requires additional data.
The regular momentum level alone determines the
presymplectic form, but not the clock forms or the
horizontal distribution.

An explicit global model exhibits the relation with
Hamilton--Jacobi theory for an Abelian group
\cite{Mukunda1978,MarmoMorandiMukunda1990}.
Let $(S,\omega_S)$ be a symplectic manifold of dimension
$2n$ and let $H_1,\dots,H_q\in C^\infty(S)$ satisfy
$\{H_a,H_b\}_{\omega_S}=0$.
Set $Y_a=X_{H_a}$ and consider
\[
\widehat P=S\times T^*\mathbb R^q,\qquad
\widehat\omega=\omega_S+\sum_{a=1}^qdt_a\wedge d\pi_a,
\qquad C_a=\pi_a+H_a.
\]
The $C_a$ Poisson-commute and have independent
differentials. Their common level $C=c$ is the graph
$\pi_a=c_a-H_a$, hence is diffeomorphic to
$S\times\mathbb R^q$.
On this graph the induced structure is
\[
\Omega=\omega_S+\sum_{a=1}^q dH_a\wedge dt_a,
\qquad \lambda_a=dt_a,\qquad
R_a=\partial_{t_a}+Y_a.
\]
Indeed, $\iota_{R_a}\Omega=0$ and
$dt_b(R_a)=\delta_{ab}$.
The characteristic equations are
$\partial x/\partial t_a=Y_a(x)$, and their
compatibility follows from $[Y_a,Y_b]=0$.
The $R_a$ generate a local $\mathbb R^q$-action,
and a global one when the $Y_a$ are complete.
Here the multi-parameter evolution is already encoded
in the Reeb fields; the functions $H_a$ should not
be confused with an additional Hamiltonian $H$ in
the formalism of Section~\ref{sec:cosymp}.

\begin{remark}\label{rem:nonabelian-origin}
The Abelian hypothesis has a precise structural role.
The cotangent-bundle description in
\cite[Sec.~III]{AlekseevskyGrabowskiMarmoMichor1994}
expresses the Liouville form through momentum mappings
and Maurer--Cartan forms.
For a left Maurer--Cartan form
$\kappa=\sum_a\kappa^a e_a$, with
$[e_a,e_b]=\sum_c c_{ab}^{\,c}e_c$, one has
\[
d\kappa^c=-\frac12\sum_{a,b}c_{ab}^{\,c}
                 \kappa^a\wedge\kappa^b.
\]
For $G=\mathbb R^q$, these components are the closed
forms $dt_a$ used in the preceding construction.

More generally, if independent fields $V_a$ satisfy
$[V_a,V_b]=\sum_c c_{ab}^{\,c}V_c$ and
$\lambda_c(V_a)=\delta_{ca}$, then
$d\lambda_c(V_a,V_b)=-c_{ab}^{\,c}$.
Closed clock forms therefore cannot be dual to a
prescribed non-Abelian frame.
Retaining such a frame requires modified structure
equations of Maurer--Cartan type along the group orbits.
This is a separate extension from the non-uniform
$q$-contact case, whose Reeb fields still commute.
\end{remark}

\subsection{From $q$-cosymplectic to uniform $q$-contact structures}

The next observation relates the two kinds of
additional data on an exact presymplectic manifold.
The horizontal distribution generally changes,
although the Reeb fields remain the same.

\begin{proposition}\label{prop:cosymp-to-contact}
Let $(M,\Omega,\vec\lambda)$ be $q$-cosymplectic,
with Reeb fields $R_1,\dots,R_q$.
Suppose that $\Omega=d\theta$ and
$\theta(R_b)=1$ for every $b$.
Then
\[
\eta_a=\theta-\sum_{c\neq a}\lambda_c,
\qquad a=1,\dots,q,
\]
define a uniform $q$-contact structure with
$d\eta_a=\Omega$ and Reeb fields $R_a$.

The normalization of the primitive can always be
achieved locally. More precisely, if
$\Omega=d\theta_0$ on a neighbourhood, then, after
shrinking it, there is a function $g$ such that
$\theta=\theta_0+dg$ satisfies $\theta(R_b)=1$
for all $b$.
\end{proposition}

\begin{proof}
We have
$\eta_a(R_b)=1-\sum_{c\neq a}\delta_{cb}=\delta_{ab}$.
Thus the $\eta_a$ are independent and their common
kernel $\xi'$ is a complement of $\mathcal R$.
Since the $\lambda_c$ are closed,
$d\eta_a=d\theta=\Omega$.
A two-form is non-degenerate on any complement of
its kernel, so $\Omega|_{\xi'}$ is non-degenerate.
These statements give the asserted uniform
$q$-contact structure and its Reeb fields.

For the local normalization, solve
$R_b(g)=1-\theta_0(R_b)$.
The compatibility conditions follow from
\[
R_c\bigl(\theta_0(R_b)\bigr)
-R_b\bigl(\theta_0(R_c)\bigr)
=d\theta_0(R_c,R_b)+\theta_0([R_c,R_b])=0.
\]
In the coordinates of Theorem~\ref{Darb},
$R_b=\partial_{t_b}$, and the Poincar\'e lemma in the
$t$ variables, with the remaining coordinates as
parameters, gives a local solution.
\end{proof}

Exactness of the clock forms is not needed in this
proposition. If $\lambda_a=d\tau_a$, its formula
becomes $\eta_a=\theta-\sum_{c\neq a}d\tau_c$.
Global existence of a normalized primitive is an
additional requirement. A global solution of the
displayed equations for $g$ is sufficient; more
generally, a primitive may be modified by a closed
one-form that need not be exact.
Thus local solvability alone gives no global
existence statement.

\subsection{Relativistic interacting particles}
\label{subsec:relativistic}

We consider the constraint formulation of
\cite[Model I]{BMMNSSZ1984}.
Let $\Gamma=T^*((\mathbb R^4)^N)$, with coordinates
$(q_a^\mu,p_{a\mu})$, where $a=1,\dots,N$ and
$\mu=0,\dots,3$.
Write $\theta=\sum_a p_{a\mu}\,dq_a^\mu$ and
$\omega=d\theta$, and use the Minkowski metric of
signature $(+,-,-,-)$.
Repeated spacetime indices are summed.
For masses $m_a>0$, consider the generalized
mass-shell constraints
\[
K_a=p_a^2-m_a^2-v_a,\qquad a=1,\dots,N.
\]
We assume that the potentials are chosen so that
$\{K_a,K_b\}_\omega=0$ and restrict to a regular
part of $M=\mathbb K^{-1}(0)$.
Strong commutativity is an assumption of this
Abelian model, not a consequence of the first-class
property alone.
The manifold $M$ has dimension $7N$, and
$\Omega=\iota^*\omega$ has rank $6N$ with
\[
\mathcal R=\ker\Omega
=\operatorname{span}\{X_{K_1}|_M,\dots,X_{K_N}|_M\}.
\]

Choose clock functions $\chi_1,\dots,\chi_N$ on a
region of $M$ such that
$A_{ab}=d\chi_a(X_{K_b}|_M)$ is invertible.
When the clocks extend to $\Gamma$, our convention
gives $A_{ab}=\{\chi_a,K_b\}_\omega|_M$.
Proposition~\ref{prop:constraints-qcos} yields an
$N$-cosymplectic structure with
$\lambda_a=d\chi_a$ and
$R_a=\sum_b(A^{-1})_{ba}X_{K_b}|_M$.
These forms specify both a normalization of the
constraint directions and a complementary
distribution.

Locally, the characteristic quotient has dimension
$6N$ and carries the reduced symplectic form.
A global reduced phase space requires the
characteristic foliation to have a smooth quotient.
Functions constant along $\mathcal R$ descend to
observables on that quotient when it exists.
The Reeb parameters describe displacements along
constraint orbits; interpreting them as individual
proper times requires information about the chosen
constraints and clocks.

\begin{example}[Free particles]\label{ex:free-relativistic}
Set $v_a=0$ and restrict to the future mass shells
$p_a^0>0$.
With $\omega=d\theta$ and
$\iota_{X_{K_a}}\omega=dK_a$, one has
$X_{K_a}=-2p_a^\mu\partial_{q_a^\mu}$.
Thus the fields
\[
R_a=\frac{p_a^\mu}{m_a^2}\,
       \frac{\partial}{\partial q_a^\mu}
\]
span $\mathcal R$.
Define $\tau_a=p_{a\mu}q_a^\mu$ on $M$.
The mass-shell equations give
$R_b(\tau_a)=\delta_{ab}$ and
$(\iota^*\theta)(R_b)=1$.
Consequently $(M,\Omega,d\tau_1,\dots,d\tau_N)$ is
globally $N$-cosymplectic, and
Proposition~\ref{prop:cosymp-to-contact} gives the
global uniform $N$-contact structure
\[
\eta_a=\iota^*\theta-\sum_{c\neq a}d\tau_c.
\]

Along an integral curve of $R_a$,
$q_a(s)=q_a(0)+s\,p_a/m_a^2$, while the other
particles remain fixed.
Since $p_a^2=m_a^2$, the proper-time increment is
$ds/m_a$, whereas $\tau_a(s)-\tau_a(0)=s$.
Thus $\tau_a$ is $m_a$ times the proper time,
up to an additive constant.
The independent Reeb flows describe the
reparametrizations of the free world lines.

The individual Liouville forms
$\theta_a=\iota^*(p_{a\mu}dq_a^\mu)$ already satisfy
$\theta_a(R_b)=\delta_{ab}$.
For $N>1$, however, $(\theta_1,\dots,\theta_N)$ is
not $N$-contact: $d\theta_a$ vanishes on all tangent
directions belonging to particles $c\neq a$, so its
kernel is larger than $\mathcal R$.
The forms
$\eta_a=\theta_a+\sum_{c\neq a}(\theta_c-d\tau_c)$
supply the required common curvature.

The cotangent lift of a proper orthochronous Lorentz
transformation preserves the future mass shells,
$\theta$, and every $\tau_a$.
It therefore preserves both structures.
The corresponding conservation laws are discussed in
Examples~\ref{ex:relativistic-noether} and
\ref{ex:relativistic-contact}.
\end{example}

For interacting particles, on any region with
admissible clocks, the restricted Liouville form is
a primitive of $\Omega$.
Proposition~\ref{prop:cosymp-to-contact} therefore
provides a uniform $N$-contact structure locally,
with the normalized constraint fields $R_a$ as Reeb
fields. This does not identify their parameters
with proper time.
Global existence requires both the clock data and
a globally normalized primitive, and is not asserted
here.

\subsection{A charge in the field of a magnetic monopole}
\label{subsec:monopole}

Let $Q=\mathbb R^3\setminus\{0\}$, let $m>0$ be the
particle mass, and assume $eg\neq0$.
Fix the magnetic two-form
$B=-eg\sin\vartheta\,d\vartheta\wedge d\varphi$.
Its flux through a sphere about the origin is
$-4\pi eg$, so it is closed but not exact.
There is consequently no global vector potential on
$Q$, and no global magnetic Lagrangian of the usual
kinetic-plus-potential form on $TQ$.
A global description is obtained by enlarging the
configuration space
\cite{BMSS1980,ZSNMMB1983}.

Let $P_1=\mathbb R_+\times SU(2)$, and define
$\pi:P_1\to Q$ by $\pi(r,s)=r\hat x$, where
$\hat x\cdot\sigma=s\sigma_3s^{-1}$ and
$\sigma_1,\sigma_2,\sigma_3$ are the Pauli matrices.
The right action $s\mapsto s e^{i\sigma_3\chi}$ makes
$P_1$ a principal $U(1)$-bundle, with generator $Z$
of period $2\pi$.
The globally defined one-form
\[
\mathcal A=ieg\,\operatorname{Tr}(\sigma_3s^{-1}ds)
\]
satisfies $\mathcal A(Z)=-2eg$ and
$d\mathcal A=\pi^*B$.

\emph{The Lagrangian description.}
For $v_u\in T_uP_1$, set
$x=\pi(u)$ and $\dot x=T_u\pi(v_u)$.
The Lagrangian is
$L(v_u)=\tfrac12m|\dot x|^2+\mathcal A_u(v_u)$
\cite{BMSS1980,ZSNMMB1983}.
It is a function on the full tangent bundle $TP_1$.
Writing $\tau:TP_1\to P_1$ and $\tau_Q:TQ\to Q$
for the projections, its Cartan form and energy are
\[
\theta_L=m\sum_{i=1}^3\dot x^i\,dx^i+\tau^*\mathcal A,
\qquad E_L=\tfrac12m|\dot x|^2.
\]
Moreover, $\omega_L=d\theta_L=(T\pi)^*\omega_B$, where
\[
\omega_B=m\sum_{i=1}^3d\dot x^i\wedge dx^i
               +\tau_Q^*B
\]
is symplectic on $TQ$.
Hence $\ker\omega_L=\ker T(T\pi)$ has rank two and is spanned
by the vertical lift $Z^V$ and the tangent lift $Z^T$ of $Z$.
For both lifts we refer to \cite{YanoIshihara1973}, where the
tangent lift is called the \emph{complete lift}.
If $\phi_t$ is the flow of $Z$ on $P_1$, then $Z^T$ generates
the flow $T\phi_t$ on $TP_1$, while $Z^V$ generates translation
in each tangent fibre, $v_u\mapsto v_u+tZ(u)$ for $v_u\in T_uP_1$.
The Cartan form satisfies $\theta_L(Z^V)=0$ and
$\theta_L(Z^T)=-2eg$, and the energy is invariant under both lifts.

The Euler--Lagrange equation
$\iota_\Gamma\omega_L=-dE_L$, together with the
second-order condition, determines $\Gamma$ up to
an arbitrary multiple of $Z^V$.
Its projection to $TQ$ is the unique vector field
$Y$ satisfying $\iota_Y\omega_B=-dE$, with
$E=\tfrac12m|\dot x|^2$.
Thus the degeneracy represents gauge freedom in the
lift, while the physical position and velocity have
a well-defined evolution.

\emph{The constraint submanifold.}
Let $\theta_0$ be the Liouville form on $T^*P_1$
and put $\omega_0=d\theta_0$.
The Legendre map is
\[
FL(v_u)=(T_u\pi)^*(m\,\dot x^\flat)+\mathcal A_u,
\]
where $\dot x^\flat$ is the Euclidean covector
associated with $\dot x$.
For $p_Z(\alpha_u)=\alpha_u(Z_u)$,
its image is the regular hypersurface
\[
M_1=\{\alpha_u\in T^*P_1:p_Z(\alpha_u)=-2eg\}.
\]
The map $FL:TP_1\to M_1$ is a surjective submersion
whose fibres are the orbits of $Z^V$.
Conversely, $\alpha_u-\mathcal A_u$ annihilates
$Z_u$, so every $\alpha_u\in M_1$ uniquely determines
$\dot x$ by the displayed formula.
Thus $M_1\cong TQ\times_QP_1$ and $\dim M_1=7$.

Let $\rho:M_1\to TQ$ be the natural projection.
We use $\theta_L$ also for the restricted Liouville
form on $M_1$, so that
$\theta_L=m\sum_i\dot x^i dx^i+\mathcal A$ and
$d\theta_L=\rho^*\omega_B$.
Here and below forms on the bundle are pulled back
to $M_1$ without changing notation.
The cotangent lift $\widehat Z$ satisfies
\[
\iota_{\widehat Z}\omega_0=-dp_Z.
\]
In our Hamiltonian convention,
$\widehat Z=-X_{p_Z}$.
It is tangent to $M_1$, is $FL$-related to $Z^T$,
and spans $\ker d\theta_L$.
We write $Z$ also for its restriction to $M_1$.

Since $\theta_L(Z)=-2eg$, the form
$\lambda=-(2eg)^{-1}\theta_L$ is contact and has
Reeb field $Z$.
Indeed, $d\lambda$ has rank six and
$\lambda(Z)=1$.
The quotient $M_1/U(1)$ is the smooth symplectic
manifold $(TQ,\omega_B)$.
The contact description retains a global potential
on the enlarged space; the nonzero flux of $B$
prevents a global potential for $\omega_B$ on $TQ$
\cite{ZSNMMB1983}.

The energy $E=\tfrac12m|\dot x|^2$ is invariant under
$Z$, and $H=E/(2eg)$ is a good contact Hamiltonian.
Its defining equations are equivalent to
\[
\iota_{X_H}d\theta_L=-dE,\qquad
\theta_L(X_H)=E.
\]
The first equation projects to the physical magnetic
dynamics, while the second fixes the gauge component.
With the standard orientation of $\mathbb R^3$,
the projected equation is
\[
m\ddot x=\frac{eg}{|x|^3}\,x\times\dot x.
\]
Thus our convention for $B$ corresponds to the
magnetic field $-g\,x/|x|^3$.
There is one physical time parameter, and the Reeb
flow changes only the gauge variable.
Rotations and their conserved quantities are
considered in Example~\ref{ex:monopole-rotations}.

\subsection{Two monopoles with irrationally related charges}
\label{subsec:two-monopoles}

Let $x_1\neq x_2$, put
$Q=\mathbb R^3\setminus\{x_1,x_2\}$, and assume
$m>0$ and $eg_1g_2\neq0$.
For $j=1,2$, let $P^{(j)}\to Q$ be the pullback of
the preceding monopole bundle under $x\mapsto x-x_j$.
Its points are represented by $(x,s_j)$ with
$s_j\sigma_3s_j^{-1}=\hat r_j\cdot\sigma$, where
$\hat r_j=(x-x_j)/|x-x_j|$.
The fibre product
$P=P^{(1)}\times_QP^{(2)}$ is a five-dimensional
principal $T^2=U(1)\times U(1)$-bundle.
Write $Z_1,Z_2$ for its generators, each of period
$2\pi$, and set
$\mathcal A_j=ieg_j\operatorname{Tr}
(\sigma_3s_j^{-1}ds_j)$.
Then
$\mathcal A_j(Z_k)=-2eg_j\delta_{jk}$ and
$d\mathcal A_j=\pi^*B_j$, where $B_j$ is the
magnetic form of the $j$-th monopole.

The global Lagrangian on $TP$ is
$L=\tfrac12m|\dot x|^2+
\mathcal A_1(v)+\mathcal A_2(v)$, with
$\dot x=T\pi(v)$ \cite{ZSNMMB1983}.
Its Cartan two-form is the pullback of
\[
\omega_B=m\sum_{i=1}^3d\dot x^i\wedge dx^i
               +\tau_Q^*(B_1+B_2).
\]
Consequently its kernel on $TP$ has rank four,
spanned by $Z_1^V,Z_2^V,Z_1^T,Z_2^T$.
The Legendre image is
\[
M_2=\{\alpha_u\in T^*P:
       \alpha_u(Z_1)=-2eg_1,\ 
       \alpha_u(Z_2)=-2eg_2\}.
\]
The two constraints are independent and
Poisson-commute.
The Legendre fibres are generated by $Z_1^V,Z_2^V$,
and $\dim M_2=8=2\cdot3+2$.

As before, every point of $M_2$ has the unique
representation
$\alpha_u=(T_u\pi)^*(m\dot x^\flat)
+\mathcal A_{1,u}+\mathcal A_{2,u}$.
Thus $M_2\cong TQ\times_QP$, and the projection
$\rho:M_2\to TQ$ is a principal $T^2$-bundle.
For the restricted Liouville form,
\[
\theta_L=m\sum_i\dot x^i\,dx^i+\mathcal A_1+\mathcal A_2,
\qquad d\theta_L=\rho^*\omega_B.
\]
Writing $Z_j$ also for the restricted cotangent
lifts, we have
$\ker d\theta_L=\mathcal R
=\operatorname{span}\{Z_1,Z_2\}$ and
$\theta_L(Z_j)=-2eg_j$.

\emph{Physical reduction and the intermediate contact quotient.}
The full torus quotient is always smooth:
$M_2/T^2\cong TQ$.
It carries the physical dynamics
$\iota_Y\omega_B=-dE$, with
$E=\tfrac12m|\dot x|^2$.
In particular,
\[
m\ddot x=e\sum_{j=1}^2
 \frac{g_j}{|x-x_j|^3}\,(x-x_j)\times\dot x.
\]
There is a single physical time parameter, and
functions of $(x,\dot x)$ are gauge-invariant
observables.

A different question is whether the lifted potential
$\theta_L$ descends to a seven-dimensional contact
manifold.
Its characteristic distribution is
\[
\ker\theta_L\cap\ker d\theta_L
=\operatorname{span}\{W\},\qquad
W=g_2Z_1-g_1Z_2.
\]
We have $\theta_L(W)=0$ and
$L_W\theta_L=0$.
If $g_1/g_2\in\mathbb Q$, this line generates a closed
circle subgroup $K\subset T^2$.
The quotient $M_2/K$ is a smooth principal circle
bundle over $TQ$, and $\theta_L$ descends to a contact
form: its exterior derivative has rank six, and the
induced one-form is nonzero on the remaining vertical
direction.

If $g_1/g_2\notin\mathbb Q$, the $W$-orbits are dense
in the fibre tori. The orbit space is not a smooth
Hausdorff manifold, so this intermediate contact
quotient is unavailable \cite{ZSNMMB1983}.
Equivalently, the magnetic periods
$-4\pi eg_1$ and $-4\pi eg_2$ cannot both be integral
multiples of one nonzero common scale, as would be
required for a single principal circle bundle with
connection curvature proportional to $\omega_B$.
This obstruction concerns a global description
retaining the potential. It does not obstruct the
full symplectic quotient $M_2/T^2$.

\emph{The non-uniform $2$-contact structure.}
The separate magnetic contributions define additional
forms on the unreduced space.

\begin{proposition}\label{prop:nonuniform}
The one-forms
\[
\lambda_j=-\frac{1}{2eg_j}
 \left(m\sum_i\dot x^i\,dx^i+\mathcal A_j\right),
\qquad j=1,2,
\]
define a non-uniform $2$-contact structure on $M_2$
with Reeb fields $Z_1,Z_2$.
Moreover, with
$\xi=\ker\lambda_1\cap\ker\lambda_2$,
the data
\[
\bigl(M_2,\,
(\lambda_1,\lambda_2)\oplus
(d\lambda_1,d\lambda_2,d\theta_L),\,
\mathcal R\oplus\xi\bigr)
\]
equip $M_2$ with a $2$-structure involving three two-forms.
\end{proposition}

\begin{proof}
The kinetic term vanishes on the gauge directions,
so $\lambda_j(Z_k)=\delta_{jk}$.
Thus the $\lambda_j$ are independent and $\xi$ is
a complement of $\mathcal R=\ker T\rho$.
Write
\[
\omega_j=m\sum_i d\dot x^i\wedge dx^i+\tau_Q^*B_j.
\]
Each $\omega_j$ is symplectic: pairing a vector in
its kernel first with all velocity directions forces
its position component to vanish, and pairing with
all position directions then forces its velocity
component to vanish.
Since
$d\lambda_j=-(2eg_j)^{-1}\rho^*\omega_j$,
we obtain $\ker d\lambda_j=\mathcal R$ and
non-degeneracy on $\xi$.
The same argument applies to
$d\theta_L=\rho^*\omega_B$.

To verify non-uniformity, suppose that
$d\lambda_1=d\lambda_2$.
Pullback by a surjective submersion is injective,
so $g_1^{-1}\omega_1=g_2^{-1}\omega_2$.
Comparison of the velocity--position terms gives
$g_1=g_2$, hence $B_1=B_2$.
This is impossible: on a small sphere about $x_1$
not enclosing $x_2$, their fluxes are $-4\pi eg_1$
and zero, respectively.
\end{proof}

Up to normalization, $\lambda_j$ is the Cartan form
for the kinetic energy and the $j$-th magnetic
contribution alone.
The physical Cartan form instead contains the
kinetic term once and both magnetic contributions.
Its derivative $d\theta_L$ governs the total motion;
it is not to be replaced by either $d\lambda_j$.
This is the reason for retaining three two-forms in
the common framework of Section~\ref{sec:prelim}.
The physical dynamics on $M_2$ is specified
presymplectically, up to the two gauge directions,
by $\iota_\Gamma d\theta_L=-dE$.
We do not apply the uniform Hamiltonian definition
to this non-uniform structure.

The next obstruction concerns all uniform structures
having the prescribed gauge fields as Reeb fields,
not only the particular forms above.

\begin{proposition}\label{prop:no-uniform}
There is no uniform $2$-contact structure
$(\eta_1,\eta_2)$ on $M_2$ whose Reeb fields are
$Z_1,Z_2$.
\end{proposition}

\begin{proof}
Suppose such a structure exists.
Then $\beta=\eta_1-\eta_2$ is closed and
$\beta(Z_1)=1$.
Choose a small sphere $S\subset Q$ about $x_1$
not enclosing $x_2$, and let
\[
N=\{(x,\dot x,u)\in TQ\times_QP:
                   x\in S,\ \dot x=0\}\subset M_2.
\]
Over $S$, the first circle bundle is the Hopf bundle,
whereas the second is trivial because it extends
over the ball bounded by $S$.
Hence $N\cong S^3\times S^1$.
An orbit $C$ of $Z_1$ is a Hopf fibre in
$S^3\times\{\mathrm{pt}\}$ and is null-homologous
in $N$.
Closedness therefore implies $\int_C\beta=0$.
On the other hand, the period of $Z_1$ is $2\pi$,
so
$\int_C\beta=\int_0^{2\pi}\beta(Z_1)\,d\chi=2\pi$,
a contradiction.
\end{proof}

\begin{remark}
The two obstructions have different meanings.
Irrationality of $g_1/g_2$ prevents the intermediate
contact quotient associated with $\theta_L$.
The topological obstruction in
Proposition~\ref{prop:no-uniform} is independent of
that ratio and prevents a uniform structure with
the prescribed Reeb fields.
It does not assert nonexistence of every possible
uniform structure on the underlying manifold.
The constructed non-uniform structure therefore lies
outside the transformation theory of
Section~\ref{sec:contact} and motivates the extension
discussed in Section~\ref{sec:conclusions}.
\end{remark}
\begin{remark}[Relation to the principal-bundle construction]
\label{rem:principal-bundle-comparison}
The Cartan forms of the monopole examples can be related
explicitly to the construction of
\cite[Sec.~V]{AlekseevskyGrabowskiMarmoMichor1994}.
Let $\pi:P\to Q$ be a principal $G$-bundle, let
$\gamma\in\Omega^1(P;\mathfrak g)$ be a principal
connection, and write $\tau_P:T^*P\to P$.
For an equivariant map $f:T^*P\to\mathfrak g^*$,
their Eqs.~(5.5)--(5.6) use the generalized Liouville form
\[
\Theta_f=\langle f,\tau_P^*\gamma\rangle,\qquad
d\Theta_f
=\langle df\wedge\tau_P^*\gamma\rangle
 +\langle f,\tau_P^*d\gamma\rangle.
\]
Here the brackets denote the natural pairing and its
extension to differential forms.
Writing
$F_\gamma=d\gamma+\tfrac12[\gamma,\gamma]$ expresses
the second term in terms of the curvature and the
Lie-algebra bracket.
Adding a horizontal Liouville term leads to the
coupled cotangent-bundle structures studied in
\cite[Sec.~V.C]{AlekseevskyGrabowskiMarmoMichor1994}.

For our examples, $G=T^q$ with $q=1$ or $2$.
Set $\mu_a=-2eg_a$ and
$\gamma^a=\mathcal A_a/\mu_a$, taking $g_1=g$ and
$\mathcal A_1=\mathcal A$ when $q=1$.
Then $\gamma^a(Z_b)=\delta_{ab}$, so the
$\gamma^a$ are the components of a principal torus
connection, and
$d\gamma^a=\pi^*(B_a/\mu_a)$.
The cotangent momenta
$f_a(\alpha_u)=\alpha_u(Z_a)$ have the fixed values
$\mu_a$ on $M_1$ or $M_2$.
With forms pulled back to the constraint manifold
and with
$\theta_{\mathrm{kin}}=m\sum_i\dot x^i\,dx^i$,
the restricted Liouville form is
\[
\theta_L=\theta_{\mathrm{kin}}+\sum_{a=1}^q\mu_a\gamma^a,
\qquad
d\theta_L=d\theta_{\mathrm{kin}}
              +\sum_{a=1}^q\mu_a\,d\gamma^a.
\]
Thus the magnetic contribution is precisely the
pairing of the fixed momentum with the curvature.

The additional forms used here are
\[
\lambda_a=\gamma^a+\mu_a^{-1}\theta_{\mathrm{kin}},
\qquad
d\lambda_a=\mu_a^{-1}
             (d\theta_{\mathrm{kin}}+d\mathcal A_a).
\]
For one monopole this is the contact normalization
$\lambda=-(2eg)^{-1}\theta_L$.
For two monopoles it gives the separate forms of
Proposition~\ref{prop:nonuniform}, whereas the physical
form $\theta_L$ contains both magnetic contributions.
This identifies the principal-bundle origin of the
physical presymplectic form and the additional data
retained by the $q$-contact description.

The bundle connection $\gamma$ generally has nonzero
curvature. It must be distinguished from the flat
horizontal projector in
Remark~\ref{rem:presymplectic-poisson}.
The principal-bundle construction therefore does not
by itself supply the closed one-forms required for
a $q$-cosymplectic structure.
\end{remark}

\section{Canonical and canonoid transformations on
$q$-cosymplectic manifolds}\label{sec:cosymp}

Throughout this section, $(M,\Omega,\vec\lambda)$ is
$q$-cosymplectic of dimension $2n+q$.
We retain the conventions of Section~\ref{sec:prelim},
in particular $\{f,g\}=X_g(f)$.
All statements about flows are local unless
completeness is explicitly assumed.
We distinguish preservation of the full data
$(\Omega,\vec\lambda)$ from preservation of the
symplectic geometry on the leaves of $\xi$.
For canonoid transformations we also specify which
parts of the Reeb data are fixed.

\subsection{Directional evolution and constants of motion}

For $H\in C^\infty(M)$ and a constant vector
$\vec a=(a_1,\dots,a_q)\in\mathbb R^q$, define
\[
E_H^{\vec a}=X_H+\sum_{i=1}^q a_iR_i.
\]
When $\vec a$ is fixed we also write $E_H$.
This is the unique field satisfying
$\iota_{E_H^{\vec a}}\Omega
=dH-\sum_iR_i(H)\lambda_i$ and
$\lambda_i(E_H^{\vec a})=a_i$.
In Darboux coordinates its integral curves obey
\[
\dot q^i=\frac{\partial H}{\partial p_i},
\qquad
\dot p_i=-\frac{\partial H}{\partial q^i},
\qquad
\dot t_j=a_j.
\]
For $q=1$ and $a_1=1$, these are the usual
non-autonomous Hamilton equations.
For several clock variables, this construction selects
a one-parameter evolution with prescribed clock
rates, within the $q$-cosymplectic framework
\cite{LeokSardonZhao}.

\begin{remark}\label{rem:directional-compatibility}
The fields $E_H^{\vec a}$ do not generally form a
commuting family. Proposition~\ref{prop:qcos-poisson}
gives
\[
[E_H^{\vec a},E_H^{\vec b}]
 =\sum_{i=1}^q(a_i-b_i)X_{R_i(H)}.
\]
For example, if $\Omega=dx\wedge dp$,
$\lambda_i=dt_i$, $q=2$, and $H=t_1p$, then
$[E_H^{e_1},E_H^{e_2}]=\partial_x$.
Thus these directional evolutions should be
distinguished from a compatible multi-parameter
evolution generated by commuting fields, such as
the Reeb fields constructed in
Section~\ref{subsec:constraint-origin}.
The entire family $E_H^{\vec a}$ commutes precisely
when $X_{R_i(H)}=0$ for all $i$.
\end{remark}

A function $f$ is a \emph{constant of motion along
the $\vec a$-direction} if
$E_H^{\vec a}(f)=0$.
Equivalently,
$\{f,H\}+\sum_i a_iR_i(f)=0$.

\begin{proposition}\label{ad}
A function $f$ is a constant of motion along every
$\vec a$-direction if and only if
$X_H(f)=0$ and $R_i(f)=0$ for every $i$.
\end{proposition}

\begin{proof}
Taking $\vec a=0$ gives $X_H(f)=0$.
Taking $\vec a=e_i$ then gives $R_i(f)=0$.
Conversely, these conditions imply
$E_H^{\vec a}(f)=0$ for every $\vec a$.
\end{proof}

This condition also makes $f$ invariant under every
field $X_H+\sum_i a_iR_i$ with smooth, rather than
constant, coefficients $a_i$.
It is stronger than conservation for one chosen
direction.

\begin{example}[Relativistic particles: observables]
\label{ex:relativistic-observables}
On a region of the constraint surface equipped with
the clocks of Section~\ref{subsec:relativistic},
take $H=0$.
Then $E_0^{\vec a}=\sum_a a_aR_a$ is tangent to
the characteristic foliation.
By Proposition~\ref{ad}, conservation for every
direction is equivalent to $R_a(f)=0$ for all $a$.
Such functions are constant along characteristic
leaves and descend to observables on a smooth reduced
phase space whenever that quotient exists
\cite{BMMNSSZ1984}.
For the free particles of
Example~\ref{ex:free-relativistic}, the clock
parameters are mass multiples of proper time.
In that case these observables are independent of
the chosen synchronization of the world lines.
No proper-time interpretation is assumed for
general interacting constraints.
\end{example}

\subsection{Canonical transformations and Noether symmetries}

\begin{definition}
A \emph{canonical transformation} is a diffeomorphism
$\psi$ satisfying $\psi^*\Omega=\Omega$ and
$\psi^*\lambda_i=\lambda_i$ for every $i$.
A vector field is \emph{canonical} if its local flow
consists of canonical transformations.
\end{definition}

\begin{remark}\label{rem:canonical-local}
A canonical transformation preserves each Reeb field:
$\psi_*R_i=R_i$.
Indeed, the pushforward satisfies the defining
contraction and normalization equations for $R_i$.
Consequently, in connected Darboux coordinate
neighbourhoods, a canonical transformation has the form
\[
\psi(q,p,t)=(Q(q,p),P(q,p),t+c),
\qquad
\sum_i dQ^i\wedge dP_i=\sum_i dq^i\wedge dp_i,
\]
where $c\in\mathbb R^q$ is constant.
The time translations cannot be discarded, since
they are themselves canonical.

These transformations preserve the full fixed
structure. A clock-dependent family of symplectic
maps can preserve the Poisson bracket without
preserving $\Omega$ as a two-form on $M$.
Such maps belong to the generalized class introduced
below; see Example~\ref{ex:gen-canonical}.
\end{remark}

The following observation isolates preservation of
the Reeb frame and the horizontal distribution from
preservation of $\Omega$.

\begin{proposition}\label{prop:q-cosymplectic-equivalence}
For a vector field $X$, the following are equivalent:
\begin{enumerate}
\item $L_X\lambda_i=0$ for every $i$ and
$\iota_{R_i}L_X\Omega=0$ for every $i$;
\item $[X,R_i]=0$ for every $i$ and
$[X,\Gamma(\xi)]\subset\Gamma(\xi)$.
\end{enumerate}
\end{proposition}

\begin{proof}
Assume the first condition.
From $\lambda_j(R_i)=\delta_{ji}$ and
$L_X\lambda_j=0$ we obtain
$\lambda_j([X,R_i])=0$.
Also,
\[
\iota_{[X,R_i]}\Omega
 =L_X(\iota_{R_i}\Omega)-\iota_{R_i}L_X\Omega=0.
\]
Thus $[X,R_i]\in\xi\cap\mathcal R=\{0\}$.
If $Y\in\Gamma(\xi)$, then
$\lambda_j([X,Y])=-(L_X\lambda_j)(Y)=0$,
so $[X,Y]\in\Gamma(\xi)$.

Conversely, assume the second condition.
The form $L_X\lambda_j$ vanishes on each $R_i$
because $[X,R_i]=0$, and on $\xi$ because
$[X,\Gamma(\xi)]\subset\Gamma(\xi)$.
Hence $L_X\lambda_j=0$.
The same contraction identity gives
$\iota_{R_i}L_X\Omega=0$.
\end{proof}

These equivalent conditions are weaker than
canonicity. For example, with $\Omega=dx\wedge dp$
and $\lambda_i=dt_i$, the field $X=x\partial_x$
satisfies both conditions but $L_X\Omega=\Omega$.
In Darboux coordinates the conditions mean that
$X=Y(q,p)+\sum_i c_i\partial_{t_i}$, where $Y$ is
horizontal and the $c_i$ are constants; no symplectic
condition on $Y$ is imposed.

\begin{lemma}\label{lem:q-cosymplectic-generators0}
A vector field $X$ is canonical if and only if
$L_X\Omega=0$ and $L_X\lambda_i=0$ for every $i$.
Locally, every such field has the form
\[
X=X_f+\sum_{i=1}^q c_iR_i,
\qquad R_i(f)=0,\qquad c_i\in\mathbb R.
\]
Moreover, for every $H$,
$[X,X_H]=X_{X(H)}$.
\end{lemma}

\begin{proof}
For the local flow $\psi_s$ of $X$,
$\frac{d}{ds}\psi_s^*\alpha=\psi_s^*L_X\alpha$.
Applying this identity to $\Omega$ and the
$\lambda_i$ proves the first assertion.

Since the $\lambda_i$ are closed,
$L_X\lambda_i=d(\lambda_i(X))=0$.
Thus $c_i=\lambda_i(X)$ is locally constant.
The one-form $\alpha=\iota_X\Omega$ is closed
and annihilates every $R_i$.
Locally $\alpha=df$, and therefore $R_i(f)=0$.
The horizontal field $X-\sum_i c_iR_i$ is then
$X_f$.

By Proposition~\ref{prop:q-cosymplectic-equivalence},
$[X,R_i]=0$.
Consequently,
\[
\begin{aligned}
\iota_{[X,X_H]}\Omega
 &=L_X(\iota_{X_H}\Omega)-\iota_{X_H}L_X\Omega\\
 &=d(X(H))-\sum_iR_i(X(H))\lambda_i.
\end{aligned}
\]
Also $\lambda_i([X,X_H])=0$, so uniqueness of
the Hamiltonian field proves the last assertion.
\end{proof}

The local normal form displays the effect of the
additional data: arbitrary components along the
presymplectic kernel are replaced by constant Reeb
translations. Globally, the existence of a function
$f$ with $df=\iota_X\Omega$ requires this closed
one-form to be exact.

\begin{definition}
An \emph{infinitesimal symmetry} of
$(M,\Omega,\vec\lambda,H)$ is a field $V$ satisfying
$L_V\Omega=0$, $L_V\lambda_i=0$ for every $i$,
and $V(H)=0$.
It is \emph{special} if also $\lambda_i(V)=0$
for every $i$.
\end{definition}

\begin{proposition}\label{prop:symmetry-flow}
A field $V$ is an infinitesimal symmetry if and only
if its local flow is canonical and preserves $H$.
Every such field satisfies
$[V,E_H^{\vec a}]=0$ for every constant $\vec a$.
\end{proposition}

\begin{proof}
The equivalence follows from
Lemma~\ref{lem:q-cosymplectic-generators0} and
$\frac{d}{ds}\phi_s^*H=\phi_s^*(V(H))$.
The same lemma gives $[V,X_H]=X_{V(H)}=0$,
and Proposition~\ref{prop:q-cosymplectic-equivalence}
gives $[V,R_i]=0$.
Thus $[V,E_H^{\vec a}]=0$.
\end{proof}

Geometric symmetry and Noether-type results in
related settings are discussed in
\cite{Carinena,AzuajeEscobar,Azuaje2,PerezAlvarez2024}.
Here the symmetry preserves the fixed forms and
the Hamiltonian separately, and the conserved
function is required to be independent of every
Reeb direction.
The following statement is the corresponding
symplectic Noether argument with these additional
compatibility conditions.

\begin{theorem}[Noether theorem]\label{thm:noether}
\begin{enumerate}
\item If $f$ is a constant of motion along every
$\vec a$-direction, then $X_f$ is a special
infinitesimal symmetry.
\item Conversely, every special infinitesimal
symmetry $V$ locally has the form $V=X_f$, where
$R_i(f)=0$ for every $i$ and $f$ is a constant of
motion along every $\vec a$-direction.
Such an $f$ can be chosen globally if and only if
the closed one-form $\iota_V\Omega$ is exact.
\end{enumerate}
\end{theorem}

\begin{proof}
For the first assertion,
Proposition~\ref{ad} gives $R_i(f)=0$ and
$X_H(f)=0$.
Hence $\iota_{X_f}\Omega=df$ and
$L_{X_f}\Omega=0$.
Horizontality and closedness of the $\lambda_i$
give $L_{X_f}\lambda_i=0$.
Finally,
$X_f(H)=\{H,f\}=-\{f,H\}=-X_H(f)=0$.

Conversely, $L_V\Omega=0$ implies
$d(\iota_V\Omega)=0$.
Choose a local primitive $f$, so
$\iota_V\Omega=df$.
Evaluating on $R_i$ gives $R_i(f)=0$.
Together with $\lambda_i(V)=0$, this yields $V=X_f$.
Furthermore,
\[
X_H(f)=df(X_H)
 =\Omega(V,X_H)
 =-dH(V)+\sum_iR_i(H)\lambda_i(V)=0.
\]
Proposition~\ref{ad} proves the conservation statement.
The same argument works globally exactly when
$\iota_V\Omega$ has a global primitive.
\end{proof}

\begin{example}
On $\mathbb R^6$ with coordinates
$(q^1,q^2,p_1,p_2,t_1,t_2)$, let
$\Omega=dq^1\wedge dp_1+dq^2\wedge dp_2$,
$\lambda_i=dt_i$, and
$H=\tfrac12(p_1^2+p_2^2)+t_1q^1$.
Then $f=p_2$ is annihilated by $X_H,R_1,R_2$.
Its Hamiltonian field $X_f=\partial_{q^2}$ is
therefore a special infinitesimal symmetry.
\end{example}

\begin{example}[Relativistic particles:
Lorentz symmetry and angular momentum]
\label{ex:relativistic-noether}
Consider the free-particle structure of
Example~\ref{ex:free-relativistic}, with $H=0$.
Write $\theta_M=\iota^*\theta$.
Let $V$ be the restriction of the cotangent lift
of an infinitesimal Lorentz transformation,
represented by a matrix $\varepsilon$ satisfying
$\varepsilon_{\mu\nu}=-\varepsilon_{\nu\mu}$.
It is tangent to the mass shells and satisfies
$L_V\theta_M=0$ and $V(\tau_a)=0$.
Thus $V$ is a special infinitesimal symmetry.

Cartan's formula gives
$\iota_V\Omega=-d(\theta_M(V))$, so the corresponding
global Hamiltonian is
\[
f=-\theta_M(V)
  =-\sum_a p_{a\mu}\varepsilon^\mu{}_\nu q_a^\nu.
\]
For
$J^{\mu\nu}=\sum_a(q_a^\mu p_a^\nu-q_a^\nu p_a^\mu)$,
this is $f=\tfrac12\varepsilon_{\mu\nu}J^{\mu\nu}$
with the stated conventions.
Since $\varepsilon$ is arbitrary, every component
of $J$ is annihilated by each $R_a$.
The total relativistic angular momentum is therefore
conserved independently of the synchronization of
the free world lines.
\end{example}

\subsection{Locally Hamiltonian vector fields and
generalized canonical transformations}

The Poisson structure of
Proposition~\ref{prop:qcos-poisson} gives a
leafwise notion of Hamiltonian vector field.
This notion need not preserve the full
$q$-cosymplectic structure.

\begin{definition}
A field $V$ is \emph{locally Hamiltonian} if
$\lambda_i(V)=0$ for every $i$ and, near every point,
there is a function $f$ such that
$\iota_V\Omega=df-\sum_iR_i(f)\lambda_i$.
It is \emph{globally Hamiltonian} if a single
globally defined $f$ has this property.
\end{definition}

Equivalently, $V$ is locally Hamiltonian if it is
horizontal and
$d(\iota_V\Omega)|_{\xi\times\xi}=0$.
Indeed, in a product Darboux chart $(x,t)$,
a horizontal one-form has the form
$\beta=\sum_a\beta_a(x,t)\,dx^a$.
When its restriction to each leaf is closed,
the Poincar\'e lemma with parameters gives a smooth
$f$ with $\partial f/\partial x^a=\beta_a$.
Thus $\beta=df-\sum_iR_i(f)\lambda_i$.
This condition is weaker than
$d(\iota_V\Omega)=0$.

\begin{definition}
A \emph{generalized canonical transformation} is a
diffeomorphism $\psi$ such that
$\psi^*\lambda_i=\lambda_i$ for every $i$ and,
locally on its domain,
\begin{equation}\label{EQ3}
\psi^*\Omega-\Omega
 =\sum_{i=1}^q dJ_i\wedge\lambda_i
\end{equation}
for smooth functions $J_i$.
Global existence of these functions is an additional
requirement when explicitly stated.
\end{definition}

\begin{proposition}\label{prop:generalized-leafwise}
For a diffeomorphism $\psi$ satisfying
$\psi^*\lambda_i=\lambda_i$, the following conditions
are equivalent:
\begin{enumerate}
\item $\psi$ is generalized canonical;
\item $(\psi^*\Omega)|_{\xi\times\xi}
       =\Omega|_{\xi\times\xi}$;
\item $\psi$ preserves the Poisson bracket.
\end{enumerate}
Locally, these maps have the form
$\psi(q,p,t)=(\Phi_t(q,p),t+c)$, where every
$\Phi_t$ is symplectic and $c$ is constant.
\end{proposition}

\begin{proof}
The first condition immediately implies the second.
Conversely, let $\Delta=\psi^*\Omega-\Omega$.
This form is closed and vanishes on
$\xi\times\xi$.
On a sufficiently small product Darboux chart choose
a one-form $\alpha$ with $d\alpha=\Delta$.
The restriction of $\alpha$ to each plaque
$t=\mathrm{const}$ is closed.
The Poincar\'e lemma with parameters therefore gives
$f$ whose differential in the leaf variables is
that restriction.
Consequently $\alpha-df=\sum_iJ_i\,dt_i$, and
$d\alpha=\sum_i dJ_i\wedge dt_i$, proving
\eqref{EQ3}.

Preservation of the $\lambda_i$ implies preservation
of $\xi$ and locally gives $t\circ\psi=t+c$.
The second condition says precisely that the induced
maps between its symplectic leaves are symplectic.
Since the Poisson tensor is the inverse of the
symplectic form on these leaves, this is equivalent
to the third condition and gives the local form.
\end{proof}

\begin{remark}\label{rem:gen-canonical}
Every canonical transformation is generalized
canonical, but a generalized canonical map need
not preserve the Reeb distribution.
In fact, \eqref{EQ3} gives
$(\psi^*\Omega)(R_k,R_l)
=R_k(J_l)-R_l(J_k)$.
For $q=2$ and $\Omega=dx\wedge dp$, the map
\[
\psi(x,p,t_1,t_2)
 =(x+t_2,p+t_1,t_1,t_2)
\]
satisfies \eqref{EQ3} with
$J_1=x+t_2$, $J_2=-p$.
Nevertheless,
$(\psi^*\Omega)(R_1,R_2)=-1$.
Thus preservation of the Poisson bracket does not
recover the full presymplectic form or its
characteristic distribution.
\end{remark}

\begin{example}\label{ex:gen-canonical}
For the same standard structure, let
$\psi(x,p,t_1,t_2)=(x,p+t_1,t_1,t_2)$.
Then $\{x,p+t_1\}=1$, but
$\psi^*\Omega=\Omega+dx\wedge dt_1\neq\Omega$.
It is generalized canonical with $J_1=x$, $J_2=0$.

The map is the time-one flow of
$V=t_1\partial_p=X_{-t_1x}$.
Thus a globally Hamiltonian field can generate
generalized canonical transformations without
being canonical.
Indeed,
$L_V\Omega=dx\wedge dt_1\neq0$.
\end{example}

\begin{theorem}\label{thm:gen-can}
Let $V$ satisfy $\lambda_i(V)=0$ for every $i$,
and let $\Psi_s$ be its local flow.
The maps $\Psi_s$ are generalized canonical if and
only if $V$ is locally Hamiltonian.
If $V=X_f$ globally, one may take
\[
J_i^{(s)}
 =-\int_0^s\Psi_u^*(R_i(f))\,du
\]
on the flow domain.
\end{theorem}

\begin{proof}
Horizontality and $d\lambda_i=0$ give
$L_V\lambda_i=0$, so $\Psi_s^*\lambda_i=\lambda_i$.
If $V=X_f$ on a neighbourhood, then
$L_V\Omega=-\sum_i d(R_i(f))\wedge\lambda_i$.
On a smaller neighbourhood where the relevant flow
segments remain in that domain,
\[
\begin{aligned}
\Psi_s^*\Omega-\Omega
 &=\int_0^s\Psi_u^*L_V\Omega\,du\\
 &=\sum_i d\left(
       -\int_0^s\Psi_u^*(R_i(f))\,du
      \right)\wedge\lambda_i.
\end{aligned}
\]
This proves the required local statement.
For longer flow segments the same conclusion follows
by composing such maps, which preserve the
symplectic structures on the leaves.
If $f$ is global, the displayed primitives are
global on the flow domain.

Conversely, Proposition~\ref{prop:generalized-leafwise}
gives
$(\Psi_s^*\Omega)|_{\xi\times\xi}
=\Omega|_{\xi\times\xi}$.
Differentiating this identity at $s=0$ yields
$d(\iota_V\Omega)|_{\xi\times\xi}=0$.
The leafwise Poincar\'e argument above proves that
$V$ is locally Hamiltonian.
This proof does not require a preliminary smooth
choice of the functions $J_i^{(s)}$.
\end{proof}

The horizontality assumption is essential:
$R_1$ generates canonical transformations but is not
locally Hamiltonian in the stated sense.

\subsection{Canonoid transformations}

We now fix $H$.
The canonoid transformations considered here preserve
the clock forms and the Reeb frame, while allowing
the symplectic form on the horizontal distribution
to change.
This specifies a restricted class of transformations
relative to the given $q$-cosymplectic data.

\begin{definition}
A possibly local diffeomorphism $F$ is
\emph{canonoid} for $(M,\Omega,\vec\lambda,H)$ if
$F^*\lambda_i=\lambda_i$ for every $i$,
$\mathcal R\subset\ker F^*\Omega$, and there is a
smooth function $K$ on its domain such that
\[
\iota_{X_H}F^*\Omega
 =dK-\sum_{i=1}^qR_i(K)\lambda_i.
\]
If this last function exists only locally, we
explicitly speak of a locally canonoid transformation.
\end{definition}

Since $F^*\Omega$ has rank $2n$, the kernel
inclusion is an equality.
Together with $F^*\lambda_i=\lambda_i$, it implies
$F_*R_i=R_i$.
Conversely, preservation of the clock forms and
each Reeb field implies the kernel condition.
In Darboux charts such maps have the form
$F(q,p,t)=(\Phi(q,p),t+c)$, with no clock dependence
in $\Phi$; $\Phi$ need not be symplectic.

Thus $(F^*\Omega,\vec\lambda)$ is
$q$-cosymplectic with the original Reeb fields,
and the definition says that $X_H$ is Hamiltonian
for this new structure.
Equivalently, in the original structure on the
target domain,
\[
F_*X_H=X_{\widetilde K},\qquad
F_*E_H^{\vec a}
 =X_{\widetilde K}+\sum_i a_iR_i,
\qquad
\widetilde K=K\circ F^{-1}.
\]
These identities follow by pushing forward the
defining contraction equations.
They explain the dynamical meaning of the definition.

Every canonical map is canonoid, with $K=H$.
Also, evaluating the defining equation on $X_H$
gives $X_H(K)=0$.
This does not imply conservation of $K$ along
every $E_H^{\vec a}$ unless also $R_i(K)=0$.

\begin{lemma}\label{lem:q-cosymplectic-generators}
Let $X$ generate a local flow $\varphi_s$ of
canonoid transformations, with new Hamiltonians
$K_s$ smooth in $(s,x)$.
Then $L_X\lambda_i=0$,
$\iota_{R_i}L_X\Omega=0$, and
\begin{equation}\label{eq:q-canonoid-cond}
\iota_{X_H}L_X\Omega
 =d\kappa-\sum_{i=1}^q R_i(\kappa)\lambda_i,
\qquad
\kappa=\left.\frac{\partial K_s}{\partial s}\right|_{s=0}.
\end{equation}
In particular, $X_H(\kappa)=0$.
\end{lemma}

\begin{proof}
Differentiate
$\varphi_s^*\lambda_i=\lambda_i$,
$\iota_{R_i}\varphi_s^*\Omega=0$, and
\[
\iota_{X_H}\varphi_s^*\Omega
 =dK_s-\sum_iR_i(K_s)\lambda_i
\]
at $s=0$.
The fields $X_H,R_i$ and the forms $\lambda_i$
are fixed in these equations, giving the stated
identities.
Evaluating \eqref{eq:q-canonoid-cond} on $X_H$
gives $X_H(\kappa)=0$.
\end{proof}

At $s=0$, the difference $K_0-H$ is a Casimir
of the Poisson bracket, namely it is constant
along $\xi$.
Locally it is a function of $t$ alone.
The same Casimir ambiguity applies to $\kappa$
in \eqref{eq:q-canonoid-cond}; it does not change
the Hamiltonian vector fields below.

\begin{theorem}\label{thm:q-cosymplectic-canonoid}
Under the hypotheses of
Lemma~\ref{lem:q-cosymplectic-generators},
$[X,R_i]=0$ for every $i$, and
\[
[X,X_H]=X_{X(H)-\kappa}.
\]
More generally, suppose that
$L_X\lambda_i=0$ and
$\iota_{R_i}L_X\Omega=0$ for every $i$.
Then \eqref{eq:q-canonoid-cond} holds for some
$\kappa$ if and only if $[X,X_H]$ is Hamiltonian.
The same equivalence holds locally.
\end{theorem}

\begin{proof}
The structural assumptions imply $[X,R_i]=0$
by Proposition~\ref{prop:q-cosymplectic-equivalence}.
First,
\begin{equation}\label{eq:q-lambda}
\lambda_i([X,X_H])
 =X(\lambda_i(X_H))-(L_X\lambda_i)(X_H)=0.
\end{equation}
Second,
\begin{equation}\label{eq:q-Omega}
\begin{aligned}
\iota_{[X,X_H]}\Omega
 &=L_X(\iota_{X_H}\Omega)-\iota_{X_H}L_X\Omega\\
 &=d(X(H))-\sum_iR_i(X(H))\lambda_i
                      -\iota_{X_H}L_X\Omega\\
 &=d(X(H)-\kappa)
       -\sum_iR_i(X(H)-\kappa)\lambda_i.
\end{aligned}
\end{equation}
This proves the first implication and the
Hamiltonian formula.

Conversely, if $[X,X_H]=X_G$, the first two lines
of \eqref{eq:q-Omega} give
\[
\iota_{X_H}L_X\Omega
 =d(X(H)-G)-\sum_iR_i(X(H)-G)\lambda_i.
\]
Thus \eqref{eq:q-canonoid-cond} holds with
$\kappa=X(H)-G$.
\end{proof}

\begin{remark}\label{rem:converse}
Call $X$ \emph{infinitesimally canonoid} if it
satisfies the structural assumptions of the theorem
and \eqref{eq:q-canonoid-cond}.
The theorem characterizes this infinitesimal
condition; it does not assert that the entire
flow is canonoid.

Indeed, the structural conditions do integrate:
$\varphi_s^*\lambda_i=\lambda_i$ and
$(\varphi_s)_*R_i=R_i$, hence
$\ker\varphi_s^*\Omega=\mathcal R$.
Put $\Omega_s=\varphi_s^*\Omega$.
On a sufficiently small product Darboux chart,
the remaining canonoid condition is equivalent to
\[
d(\iota_{X_H}\Omega_s)|_{\xi\times\xi}=0
\quad\text{for every sufficiently small }s.
\]
The equivalence follows from the Poincar\'e lemma
on the leaves, with the clock variables and $s$
as parameters.
Global Hamiltonians require global primitives
of the same horizontal equations.

The infinitesimal condition controls only the
first derivative at $s=0$ of this family of
equations.
The next example shows that the second derivative
can already obstruct integration.
\end{remark}

\begin{example}[An infinitesimal condition that does
not integrate]\label{ex:infinitesimal-not-canonoid}
Let $M=\mathbb R^{2+q}$ with
$\Omega=dx\wedge dp$, $\lambda_i=dt_i$, and $H=p$.
Then $X_H=\partial_x$.
Consider
\[
X=xp\,\partial_x+x\,\partial_p.
\]
It preserves the clock forms and commutes with
every $R_i=\partial_{t_i}$.
Moreover,
$L_X\Omega=p\,\Omega$, so
\[
\iota_{X_H}L_X\Omega=p\,dp=d(p^2/2).
\]
Thus $X$ is infinitesimally canonoid with
$\kappa=p^2/2$.
Correspondingly,
\[
[X,X_H]=-p\,\partial_x-\partial_p
 =X_{\,x-p^2/2},
\qquad X(H)-\kappa=x-p^2/2.
\]

Nevertheless, its local flow is not a flow of
canonoid transformations.
Since $X$ does not involve the clock variables,
write $\varphi_s^*\Omega=A_s(x,p)\,dx\wedge dp$.
We have
$L_X^2\Omega=(x+p^2)\Omega$, and hence, locally
with the remainder also controlled after one
spatial derivative,
\[
A_s=1+sp+\frac{s^2}{2}(x+p^2)+O(s^3).
\]
Therefore
\[
\begin{aligned}
\iota_{X_H}\varphi_s^*\Omega&=A_s\,dp,\\
d(\iota_{X_H}\varphi_s^*\Omega)|_{\xi\times\xi}
 &=\partial_xA_s\,dx\wedge dp\\
 &=\left(\frac{s^2}{2}+O(s^3)\right)dx\wedge dp.
\end{aligned}
\]
At $(x,p)=(0,0)$ this is nonzero for sufficiently
small $s\neq0$.
Thus no local function $K_s$ can satisfy the
canonoid equation near that point.
The counterexample works for every $q\geq1$.
\end{example}

There are, however, useful sufficient conditions
that can be checked for the whole flow.
They include the following versions of dynamical
symmetries and constant rescalings
\cite{Azuaje2}.

\begin{proposition}\label{prop:qcos-canonoid-sufficient}
Suppose that $L_X\lambda_i=0$ and
$\iota_{R_i}L_X\Omega=0$ for every $i$.
Either of the following conditions ensures that
the local flow $\varphi_s$ is canonoid:
\begin{enumerate}
\item $[X,X_H]=0$, in which case
$K_s=\varphi_s^*H$;
\item $L_X\Omega=c\Omega$ for a constant $c$,
in which case $K_s=e^{cs}H$.
\end{enumerate}
The second condition works for every Hamiltonian $H$.
\end{proposition}

\begin{proof}
The structural conditions preserve the clock forms
and the Reeb fields along the flow.
In the first case, $X_H$ is also preserved, so
\[
\begin{aligned}
\iota_{X_H}\varphi_s^*\Omega
 &=\varphi_s^*(\iota_{X_H}\Omega)\\
 &=d(\varphi_s^*H)
       -\sum_iR_i(\varphi_s^*H)\lambda_i.
\end{aligned}
\]
In the second case,
$\varphi_s^*\Omega=e^{cs}\Omega$, and therefore
\[
\iota_{X_H}\varphi_s^*\Omega
 =d(e^{cs}H)-\sum_iR_i(e^{cs}H)\lambda_i.
\]
These explicit families prove canonoidness.
\end{proof}

\begin{example}[A canonoid, non-canonical flow with
explicit clock dependence in the Hamiltonian]
\label{ex:canonoid-cosymp}
Let $\Omega=dx\wedge dp$, $\lambda_i=dt_i$ for
$i=1,2$, and
$H=\tfrac12p^2+(t_1+t_2)x$.
Then
$X_H=p\partial_x-(t_1+t_2)\partial_p$.
The field $X=p\partial_p$ has flow
$\varphi_s(x,p,t)=(x,e^sp,t)$ and satisfies
$L_X\Omega=\Omega$.
Thus $\varphi_s$ is canonoid with $K_s=e^sH$,
although it is not canonical for $s\neq0$.

Here $\kappa=H$, so
$X(H)-\kappa=\tfrac12p^2-(t_1+t_2)x$.
Directly,
\[
[X,X_H]
 =p\partial_x+(t_1+t_2)\partial_p
 =X_{\,\frac12p^2-(t_1+t_2)x}.
\]
More generally, on the standard product model,
$X=\sum_i p_i\partial_{p_i}$ generates such a
canonoid flow for every $H$, with
$[X,X_H]=X_{X(H)-H}$.
The rescaling field need not extend globally to
an arbitrary $q$-cosymplectic manifold.
\end{example}

\section{Canonical and canonoid transformations on
uniform $q$-contact manifolds}\label{sec:contact}

Throughout this section, $(M,\vec\lambda)$ is a
uniform $q$-contact manifold, with
$\Omega=d\lambda_1=\cdots=d\lambda_q$.
Its Reeb fields are denoted by $R_1,\dots,R_q$,
and $E=\sum_iR_i$.
We use the Hamiltonian fields and Jacobi bracket
of Section~\ref{sec:prelim}.
All flow statements are local unless completeness
is assumed.

We study transformations relative to the full
ordered family $\vec\lambda$.
The distinction from the underlying Jacobi
structure is relevant here: strict preservation
of every $\lambda_i$ imposes conditions on the
transverse data as well as on the contact
structures of the leaves.

\subsection{Canonical transformations and symmetries}

\begin{definition}
A \emph{canonical transformation} is a possibly
local diffeomorphism $F$ such that
$F^*\lambda_i=\lambda_i$ for every $i$.
A field $X$ is \emph{canonical} if its local flow
consists of canonical transformations.
\end{definition}

The flow identity
$\frac{d}{ds}\phi_s^*\lambda_i=\phi_s^*L_X\lambda_i$
shows that $X$ is canonical precisely when
$L_X\lambda_i=0$ for every $i$.
These fields form a Lie subalgebra, since
$L_{[X,Y]}\lambda_i
=L_XL_Y\lambda_i-L_YL_X\lambda_i$.
They do not form a $C^\infty(M)$-module:
$L_{fX}\lambda_i=fL_X\lambda_i+\lambda_i(X)\,df$.
For example, $R_i$ is canonical, whereas
$L_{fR_i}\lambda_i=df$.

\begin{proposition}\label{PP}
For a vector field $X$, the following are equivalent:
\begin{enumerate}
\item $L_X\lambda_i=0$ for every $i$;
\item $[X,R_i]=0$ for every $i$ and
$[X,\Gamma(\xi)]\subset\Gamma(\xi)$.
\end{enumerate}
\end{proposition}

\begin{proof}
Assume the first condition.
Differentiating $\lambda_j(R_i)=\delta_{ji}$
gives $\lambda_j([X,R_i])=0$.
Also $L_X\Omega=dL_X\lambda_1=0$, so
\[
\iota_{[X,R_i]}\Omega
=L_X(\iota_{R_i}\Omega)-\iota_{R_i}L_X\Omega=0.
\]
Hence $[X,R_i]\in\xi\cap\mathcal R=\{0\}$.
For $Y\in\Gamma(\xi)$,
$\lambda_i([X,Y])=-(L_X\lambda_i)(Y)=0$,
which proves preservation of $\xi$.

Conversely, the second condition implies that
$L_X\lambda_j$ vanishes on every $R_i$ and on
every field with values in $\xi$.
Since $TM=\mathcal R\oplus\xi$, it follows that
$L_X\lambda_j=0$.
\end{proof}

Unlike the analogous conditions in
Proposition~\ref{prop:q-cosymplectic-equivalence},
these conditions already imply preservation of
the two-form, because $\Omega=d\lambda_i$.

The following description makes the restrictions
on canonical fields more explicit.

\begin{proposition}\label{prop:contact-canonical-fields}
On each connected component, every canonical
field $V$ admits the unique decomposition
\[
V=X_h+\sum_{i=2}^q c_iR_i,
\qquad
h=-\lambda_1(V),\qquad
R_j(h)=0,\qquad c_i\in\mathbb R.
\]
The sum is empty when $q=1$.
Conversely, every field of this form is canonical.
For every canonical field $V$ and every
Hamiltonian $H$,
$[V,X_H]=X_{V(H)}$.
\end{proposition}

\begin{proof}
Put $f_i=\lambda_i(V)$.
Cartan's formula gives
$df_i=-\iota_V\Omega$ for every $i$.
Therefore $R_j(f_i)=0$ and
$d(f_i-f_1)=0$.
On a connected component, write
$c_i=f_i-f_1$, with $c_1=0$, and set $h=-f_1$.
Then $R_j(h)=0$ and $\iota_V\Omega=dh$.

The field $X_h$ satisfies
$\iota_{X_h}\Omega=dh$ and
$\lambda_i(X_h)=-h=f_1$.
Thus $V-X_h$ belongs to $\mathcal R$, and its
contractions with the $\lambda_i$ are the constants
$c_i$. This proves the decomposition and uniqueness.
Conversely, $R_j(h)=0$ implies
$L_{X_h}\lambda_i=0$, and each $R_i$ is canonical.

Finally, Proposition~\ref{PP} gives
$[V,R_i]=0$, while $L_V\Omega=0$.
Consequently,
\[
\begin{aligned}
\lambda_i([V,X_H])&=-V(H),\\
\iota_{[V,X_H]}\Omega
 &=d(V(H))-\sum_j R_j(V(H))\lambda_j.
\end{aligned}
\]
These are the defining equations for $X_{V(H)}$.
\end{proof}

In particular, a canonical field is Hamiltonian
in the sense of Section~\ref{sec:prelim} precisely
when all its contractions $\lambda_i(V)$ coincide.
For $q>1$, the remaining constant Reeb components
must also be retained.

\begin{definition}
An \emph{infinitesimal symmetry} of
$(M,\vec\lambda,H)$ is a canonical field $V$
such that $V(H)=0$.
We use \emph{dynamical symmetry} for a field
satisfying $[V,X_H]=0$, without requiring
preservation of the defining forms.
\end{definition}

\begin{remark}\label{rem:symmetries-commute}
Infinitesimal symmetries form a Lie subalgebra
of the canonical fields.
By Proposition~\ref{prop:contact-canonical-fields},
every infinitesimal symmetry is a dynamical
symmetry. The converse need not hold.
\end{remark}

\begin{proposition}
A field $V$ is an infinitesimal symmetry if and
only if its local flow is canonical and preserves
$H$.
\end{proposition}

\begin{proof}
Apply
$\frac{d}{ds}\phi_s^*\alpha=\phi_s^*L_V\alpha$
to each $\lambda_i$ and to $H$.
Vanishing of the Lie derivatives makes the
pullbacks constant in $s$.
Conversely, differentiation of the invariance
identities at $s=0$ gives the defining conditions.
\end{proof}

\subsection{Dissipated $q$-tuples}

Dissipated quantities and their relation to
symmetries in ordinary contact mechanics are
discussed in \cite{PerezAlvarez2024,Azuaje2}.
Here the identity
$L_{X_H}\lambda_i=-\sum_jR_j(H)\lambda_j$
leads to a coupled law for the contractions with
the full family of one-forms.

\begin{definition}
A \emph{dissipated $q$-tuple} is a tuple
$(f_1,\dots,f_q)$ satisfying
\[
X_H(f_i)=-\sum_{j=1}^qR_j(H)f_j,
\qquad i=1,\dots,q.
\]
\end{definition}

For any such tuple, $X_H(f_i-f_j)=0$.
Thus the differences are first integrals, although
they need not be constant functions for an arbitrary
dissipated tuple.

More generally, for any vector field $V$,
\[
X_H(\lambda_i(V))
=-\sum_jR_j(H)\lambda_j(V)
  +\lambda_i([X_H,V]).
\]
This follows from the formula for
$L_{X_H}\lambda_i$.
In particular, every dynamical symmetry gives a
dissipated tuple.
Strict infinitesimal symmetries impose the
additional restrictions in the next theorem.

\begin{theorem}\label{TI}
Let $V$ be an infinitesimal symmetry and set
$f_i=\lambda_i(V)$.
Then, for every $i,k$,
\[
X_H(f_i)=-\sum_{j=1}^qR_j(H)f_j,
\qquad R_k(f_i)=0,
\qquad d(f_i-f_k)=0.
\]
Consequently:
\begin{enumerate}
\item $(f_1,\dots,f_q)$ is a dissipated tuple;
\item its differences are locally constant;
\item every $f_i$ is a first integral if and only if
$\sum_jR_j(H)f_j=0$.
\end{enumerate}
\end{theorem}

\begin{proof}
Since $V$ is a dynamical symmetry, the general
contraction identity above gives the dissipation law.
Canonicity gives $df_i=-\iota_V\Omega$ for every $i$.
Evaluation on $R_k$ yields $R_k(f_i)=0$,
and subtraction gives $d(f_i-f_k)=0$.
The last assertion follows directly from the
common right-hand side of the dissipation equations.
\end{proof}

Thus a fixed strict symmetry contributes at most
one functionally independent nonconstant function
among its $q$ contractions.
Its constant differences should not be counted
as additional independent conservation laws.
In the decomposition of
Proposition~\ref{prop:contact-canonical-fields},
$f_i=-h+c_i$, with $c_1=0$, and the common
dissipation equation is equivalently
\[
X_H(h)=-hE(H)+\sum_{i=2}^q c_iR_i(H).
\]

\begin{remark}
The infinitesimal symmetries for which all
$\lambda_i(V)$ are first integrals form a Lie
ideal in the algebra of infinitesimal symmetries.
Indeed, write
$S_V=\sum_iR_i(H)\lambda_i(V)$.
For another infinitesimal symmetry $W$,
$[W,R_i]=0$, $W(H)=0$, and $L_W\lambda_i=0$
give $W(S_V)=S_{[W,V]}$.
Hence $S_V=0$ implies $S_{[W,V]}=0$.
\end{remark}

\begin{corollary}
For a good Hamiltonian system, every infinitesimal
symmetry $V$ gives first integrals
$\lambda_1(V),\dots,\lambda_q(V)$.
They are invariant under all Reeb flows and differ
only by locally constant functions.
\end{corollary}

For comparison, the scalar dissipation law
associated with the Jacobi structure is
$X_H(f)=-E(H)f$.
It satisfies the familiar Hamiltonian symmetry
correspondence:
\[
X_H(f)=-E(H)f
\quad\Longleftrightarrow\quad
\{f,H\}=0
\quad\Longleftrightarrow\quad
[X_f,X_H]=0.
\]
The first equivalence uses
$\{f,H\}=X_H(f)+fE(H)$.
The second follows from
$[X_f,X_H]=-X_{\{f,H\}}$ and injectivity of
$f\mapsto X_f$.
Thus scalar dissipated quantities correspond
to Hamiltonian dynamical symmetries, which need
not preserve each $\lambda_i$ strictly.

\begin{example}[Linear damping]
Let $n=1$, $q=2$, and
$\lambda_i=dz_i-p\,dx$.
For $H=\tfrac12p^2+\gamma z_1$, with $\gamma>0$,
the equation for $p$ is $\dot p=-\gamma p$.
The field $V=\partial_x$ is an infinitesimal
symmetry, and
$\lambda_1(V)=\lambda_2(V)=-p$.
Theorem~\ref{TI} gives
$X_H(-p)=\gamma p$, hence
$p(t)=p(0)e^{-\gamma t}$.
The two contractions express the same damping
law, rather than two independent dissipated
quantities.
\end{example}

\begin{example}[A dynamical symmetry with a
nonconstant difference]
On the same structure, let $H=p^2/2$ and
$u=z_2-z_1$.
Then
$X_H=p\partial_x+\tfrac12p^2(R_1+R_2)$,
$X_H(u)=0$, and $[X_H,R_2]=0$.
Consequently $V=uR_2$ is a dynamical symmetry.
Its contractions are $(0,u)$, a dissipated tuple
with nonconstant difference.
However, $L_V\lambda_2=du\neq0$.
This illustrates why local constancy of the
differences in Theorem~\ref{TI} uses strict
preservation of the defining forms.
\end{example}

\begin{example}[Charge--monopole system: rotations]
\label{ex:monopole-rotations}
Consider the contact manifold $(M_1,\lambda)$
of Section~\ref{subsec:monopole}, with
$\lambda=-\theta_L/(2eg)$ and good Hamiltonian
$H=E/(2eg)$.
For $\alpha=1,2,3$, let $V_\alpha$ be the
restriction of the cotangent lift of the
left-action generator
$s\mapsto e^{i\sigma_\alpha\epsilon/2}s$.
This action commutes with the gauge action and
preserves $\mathcal A$.
Its lift therefore preserves $M_1$ and $\theta_L$.
Since the kinetic energy is rotation invariant,
$V_\alpha$ is an infinitesimal symmetry
\cite{BMSS1980}.

With the conventions of
Section~\ref{subsec:monopole}, the Pauli identity
$[\sigma_\alpha,\sigma_\beta]
=2i\epsilon_{\alpha\beta\gamma}\sigma_\gamma$
gives $V_\alpha(x)=-e_\alpha\times x$.
Also,
\[
\mathcal A(V_\alpha)=-eg\,\hat x_\alpha,
\qquad
m\dot x\cdot V_\alpha(x)
=-(x\times m\dot x)_\alpha.
\]
Thus
\[
\lambda(V_\alpha)=\frac{J_\alpha}{2eg},
\qquad
J=x\times m\dot x+eg\,\hat x.
\]
By Theorem~\ref{TI}, these functions are conserved.
They give the usual angular momentum with its
Poincar\'e monopole term \cite{BMSS1980}.

The sign is consistent with the equation of
motion in Section~\ref{subsec:monopole}:
\[
\frac{d}{dt}(x\times m\dot x)
 =\frac{eg}{|x|^3}x\times(x\times\dot x)
 =-eg\,\frac{d\hat x}{dt}.
\]
Hence $dJ/dt=0$ directly.
\end{example}

\begin{example}[Relativistic particles:
conservation along the Reeb flows]
\label{ex:relativistic-contact}
Consider the free uniform $N$-contact structure
of Example~\ref{ex:free-relativistic}.
Write $\theta_M=\iota^*\theta$, and let $V$
be an infinitesimal Lorentz transformation.
Since $L_V\theta_M=0$ and $V(\tau_c)=0$,
each $\eta_a$ is preserved and
$\eta_a(V)=\theta_M(V)$.
For the conventions of
Example~\ref{ex:relativistic-noether},
\[
\eta_a(V)
=-\tfrac12\varepsilon_{\mu\nu}J^{\mu\nu},
\qquad a=1,\dots,N.
\]
Canonicity alone gives
$R_k(\eta_a(V))=0$ for every $k$, as in
Proposition~\ref{prop:contact-canonical-fields}.
Thus the angular momentum is conserved along
each independent free-particle proper-time flow.

This statement concerns the Reeb dynamics.
If one additionally specifies a Lorentz-invariant
good Hamiltonian, for example
$H=P_\mu P^\mu$ with $P=\sum_a p_a$,
then $V$ is an infinitesimal symmetry of that
Hamiltonian system.
The same functions are then also first integrals
of $X_H$.
The choice $H=0$ gives $X_H=0$ and is not needed
to establish conservation along the Reeb fields.
\end{example}

\subsection{Canonoid transformations}

Every diffeomorphism $F$ pulls the given structure
back to a uniform $q$-contact structure.
Its Reeb fields are
$F^*R_j=(F^{-1})_*R_j$, with the vector-field
pullback understood on the domain of $F$.
In contrast to the class used in
Section~\ref{sec:cosymp}, we do not require these
fields to equal the original $R_j$.

\begin{definition}
A possibly local diffeomorphism $F$ is
\emph{canonoid} for $(M,\vec\lambda,H)$ if there
is a smooth function $K$ on its domain such that
\[
\iota_{X_H}F^*\lambda_i=-K,\qquad
\iota_{X_H}dF^*\lambda_i
 =dK-\sum_{j=1}^q(F^*R_j)(K)\,F^*\lambda_j
\]
for every $i=1,\dots,q$.
\end{definition}

The same function $K$ must satisfy all the
contraction equations.
In particular,
$(F^*(\lambda_i-\lambda_k))(X_H)=0$ for every
$i,k$, and $K$ is uniquely determined whenever
these equations hold.
Pushing forward the defining equations gives
the equivalent statement
$F_*X_H=X_{\widetilde K}$, where
$\widetilde K=K\circ F^{-1}$.

Thus, if $(Q,P,Z)$ are Darboux coordinates on
the target, the transformed equations are
\[
\begin{aligned}
\dot Q^a
 &=\frac{\partial\widetilde K}{\partial P_a},\\
\dot P_a
 &=-\frac{\partial\widetilde K}{\partial Q^a}
   -P_a\sum_{j=1}^q
           \frac{\partial\widetilde K}{\partial Z_j},\\
\dot Z_i
 &=\sum_{a=1}^nP_a
           \frac{\partial\widetilde K}{\partial P_a}
   -\widetilde K.
\end{aligned}
\]
The distinction between $K$ on the source and
$\widetilde K$ on the target avoids identifying
two different coordinate expressions.
A canonical transformation is canonoid with $K=H$.

\begin{theorem}\label{thm:contact-canonoid}
Let $X$ generate a local flow $\varphi_s$ of
canonoid transformations, and let $K_s$ be the
corresponding Hamiltonians.
Then $K_0=H$, the functions $K_s$ depend smoothly
on $(s,x)$, and, with
$\kappa=\left.\partial_sK_s\right|_{s=0}$,
\[
\kappa=-(L_X\lambda_i)(X_H)
\quad\text{for every }i,
\qquad
[X,X_H]=X_{X(H)-\kappa}.
\]
\end{theorem}

\begin{proof}
Set $Y=X_H$.
The canonoid equations are
\begin{equation}\label{eq:contact-canonoid-s}
\begin{aligned}
(\varphi_s^*\lambda_i)(Y)&=-K_s,\\
\iota_Yd\varphi_s^*\lambda_i
 &=dK_s-\sum_j(\varphi_s^*R_j)(K_s)
                         \,\varphi_s^*\lambda_j.
\end{aligned}
\end{equation}
The first identity determines $K_s$ uniquely and
smoothly, and gives $K_0=H$.
Its derivative at zero yields
\begin{equation}\label{eq:contact-kappa}
\kappa=-(L_X\lambda_i)(Y),
\qquad i=1,\dots,q.
\end{equation}
Since
$\left.\partial_s(\varphi_s^*R_j)\right|_{s=0}
=[X,R_j]$, differentiating the second identity
gives
\begin{equation}\label{eq:contact-diff}
\iota_YdL_X\lambda_i
 =d\kappa
 -\sum_j\bigl([X,R_j](H)+R_j(\kappa)\bigr)\lambda_j
 -\sum_jR_j(H)L_X\lambda_j.
\end{equation}
Together with \eqref{eq:contact-kappa}, Cartan's
formula equivalently gives
\begin{equation}\label{eq:contact-diff2}
L_YL_X\lambda_i
 =-\sum_j\bigl([X,R_j](H)+R_j(\kappa)\bigr)\lambda_j
  -\sum_jR_j(H)L_X\lambda_j.
\end{equation}

Now
$\lambda_i([X,Y])
=X(\lambda_i(Y))-(L_X\lambda_i)(Y)
=-X(H)+\kappa$.
Moreover, using \eqref{eq:contact-diff} and
$d\lambda_i=\Omega$,
\[
\begin{aligned}
\iota_{[X,Y]}\Omega
 &=L_X(\iota_Y\Omega)-\iota_YL_X\Omega\\
 &=d(X(H))-\sum_jX(R_j(H))\lambda_j
       -\sum_jR_j(H)L_X\lambda_j
       -\iota_YdL_X\lambda_i\\
 &=d(X(H)-\kappa)
       -\sum_jR_j(X(H)-\kappa)\lambda_j.
\end{aligned}
\]
Here $X(R_j(H))-[X,R_j](H)=R_j(X(H))$.
The contraction equations therefore identify
$[X,Y]$ with $X_{X(H)-\kappa}$.
\end{proof}

The Hamiltonian formula has the same form as in
Section~\ref{sec:cosymp}.
Here, however, $\kappa$ is fixed by the contractions
with the one-forms, and the Hamiltonian assignment
is injective.
There is no Casimir ambiguity.

\begin{remark}\label{rem:contact-infinitesimal}
Call $X$ \emph{infinitesimally canonoid} if there
is a function $\kappa$ satisfying
\eqref{eq:contact-kappa} and
\eqref{eq:contact-diff}.
These first-order conditions are equivalent to
$[X,X_H]$ being Hamiltonian.

Indeed, the theorem's calculation proves one
implication.
Conversely, suppose $[X,X_H]=X_G$ and put
$\kappa=X(H)-G$.
Contraction with $\lambda_i$ gives
$-(L_X\lambda_i)(X_H)=\kappa$ for every $i$.
Also,
\[
\begin{aligned}
\iota_{X_H}dL_X\lambda_i
 &=L_X(\iota_{X_H}\Omega)-\iota_{X_G}\Omega\\
 &=d\kappa
   -\sum_j\bigl([X,R_j](H)+R_j(\kappa)\bigr)\lambda_j
   -\sum_jR_j(H)L_X\lambda_j,
\end{aligned}
\]
which is \eqref{eq:contact-diff}.
This characterizes the infinitesimal condition,
not the canonoid property of the entire flow.
\end{remark}

\begin{example}[An infinitesimal condition that does
not integrate]\label{ex:contact-infinitesimal-counterexample}
Let $M=\mathbb R^{2+q}$, with
$\lambda_i=dz_i-p\,dx$, and take $H=p$.
Then $X_H=\partial_x$.
Consider
\[
X=z_1\partial_p+x\sum_{j=1}^q\partial_{z_j}.
\]
We have
$L_X\lambda_i=(1-z_1)\,dx$ for every $i$,
$\kappa=z_1-1$, and
\[
[X,X_H]=-\sum_j\partial_{z_j}=X_1,
\qquad X(H)-\kappa=1.
\]
Thus $X$ is infinitesimally canonoid.

Its flow is globally explicit:
\[
\varphi_s(x,p,z_1,\dots,z_q)
 =\left(
 x,\ p+sz_1+\tfrac12s^2x,\
 z_1+sx,\dots,z_q+sx
 \right).
\]
Consequently,
\[
\varphi_s^*\lambda_i
 =\lambda_i+
   \left(s-sz_1-\tfrac12s^2x\right)dx.
\]
The first canonoid equation forces the unique
candidate
$K_s=p-s+sz_1+\tfrac12s^2x$.
The pulled-back Reeb fields are
$\varphi_s^*R_1=\partial_{z_1}-s\partial_p$
and $\varphi_s^*R_j=\partial_{z_j}$ for $j\geq2$.
All of them annihilate $K_s$.
However,
\[
\iota_{X_H}d\varphi_s^*\lambda_i
 =dp+s\,dz_1,
\qquad
dK_s=dp+s\,dz_1+\tfrac12s^2dx.
\]
The second canonoid equation therefore fails for
every $s\neq0$.
This provides a finite-flow obstruction for every
$q\geq1$, even though the original Hamiltonian
is good.
\end{example}

\subsection{Sufficient conditions for canonoid flows}

The preceding example requires a distinction
between first-order tests and hypotheses that
control the whole flow.
Two useful sufficient conditions follow.

\begin{proposition}\label{prop:contact-dynamical-canonoid}
If $[X,X_H]=0$, the local flow $\varphi_s$ of $X$
is canonoid, with $K_s=\varphi_s^*H$.
\end{proposition}

\begin{proof}
The flow preserves $Y=X_H$.
Hence
$(\varphi_s^*\lambda_i)(Y)
=\varphi_s^*(\lambda_i(Y))=-\varphi_s^*H$.
Naturality of contraction and differentiation
also gives
\[
\begin{aligned}
\iota_Yd\varphi_s^*\lambda_i
 &=\varphi_s^*(\iota_Y\Omega)\\
 &=d(\varphi_s^*H)
   -\sum_j(\varphi_s^*R_j)(\varphi_s^*H)
                         \,\varphi_s^*\lambda_j.
\end{aligned}
\]
These are the canonoid equations.
\end{proof}

\begin{proposition}[Common constant rescalings]
\label{prop:conformal}
Suppose $L_X\lambda_i=c\lambda_i$ for every $i$,
where $c\in\mathbb R$ is constant.
Then the local flow of $X$ is canonoid for every
Hamiltonian $H$, with
\[
K_s=e^{cs}H,\qquad
[X,X_H]=X_{X(H)-cH}.
\]
Moreover, when $q\geq2$, if
$L_X\lambda_i=\mu\lambda_i$ for every $i$
with a common smooth function $\mu$, then
$\mu$ is constant on each connected component.
\end{proposition}

\begin{proof}
For constant $c$, the flow satisfies
$\varphi_s^*\lambda_i=e^{cs}\lambda_i$ and
$d\varphi_s^*\lambda_i=e^{cs}\Omega$.
The defining Reeb equations give
$\varphi_s^*R_j=e^{-cs}R_j$.
Therefore
$(\varphi_s^*\lambda_i)(X_H)=-e^{cs}H$, and
\[
\iota_{X_H}d\varphi_s^*\lambda_i
 =d(e^{cs}H)
  -\sum_j(e^{-cs}R_j)(e^{cs}H)\,e^{cs}\lambda_j.
\]
This proves canonoidness with $K_s=e^{cs}H$.
Theorem~\ref{thm:contact-canonoid}, with
$\kappa=cH$, gives the commutator formula.

Now suppose $q\geq2$ and
$L_X\lambda_i=\mu\lambda_i$.
Set $f_i=\lambda_i(X)$.
Cartan's formula, evaluated on $R_k$, gives
$R_k(f_i)=\mu\delta_{ik}$.
For $k\neq i$, commutativity of the Reeb fields
therefore yields
$R_k(\mu)=R_kR_i(f_i)=R_iR_k(f_i)=0$.
Choosing $i\neq k$ proves $R_k(\mu)=0$ for every $k$.

Differentiating the common rescaling equations gives
\[
L_X\Omega=d\mu\wedge\lambda_i+\mu\Omega
\quad\text{for every }i.
\]
Thus $d\mu\wedge(\lambda_i-\lambda_j)=0$.
For $Y\in\xi$ and $i\neq j$, evaluation on
$(Y,R_i)$ gives $d\mu(Y)=0$.
Since $TM=\xi\oplus\mathcal R$, we conclude that
$d\mu=0$.
\end{proof}

The last assertion is a restriction imposed by
simultaneously rescaling at least two defining forms
of a uniform structure.
For $q=1$, a common conformal rate is simply the
ordinary contact conformal rate and need not be
constant.
The constant-rescaling case above is the direct
analogue of the scaling constructions in
\cite{Azuaje2}.

\begin{example}
In the Darboux coordinates of
Proposition~\ref{prop:darboux-contact}, the field
\[
X=\sum_a p_a\partial_{p_a}
  +\sum_j z_j\partial_{z_j}
\]
satisfies $L_X\lambda_i=\lambda_i$.
Its flow is
$\varphi_s(q,p,z)=(q,e^sp,e^sz)$, so it is
canonoid for every $H$, with $K_s=e^sH$.

For $n=1$ and $H=p^2/2$,
$X_H=p\partial_x+\tfrac12p^2\sum_j\partial_{z_j}$.
Since $X(H)-H=H$, direct computation gives
$[X,X_H]=X_H$, as predicted.
This construction is local on a general manifold;
a global rescaling field is an additional
geometric requirement.
\end{example}

The results of this section use uniformity
throughout. In particular, the common contraction
equations, the constant differences for strict
symmetries, and the rescaling restriction depend
on $d\lambda_1=\cdots=d\lambda_q$.
The two-monopole structure of
Proposition~\ref{prop:nonuniform} requires a
separate non-uniform theory.

\section{Conclusions and open problems}\label{sec:conclusions}

We have studied canonical and canonoid transformations on
$q$-cosymplectic and uniform $q$-contact manifolds, with particular
attention to their origin in constrained Hamiltonian mechanics.
The underlying presymplectic geometry determines a characteristic
distribution, while the additional one-forms specify a complementary
distribution and a distinguished frame of characteristic fields.
Our transformation theory is formulated relative to these additional
data. Regular first-class constraint surfaces admit local
$q$-cosymplectic completions after a choice of transverse clock
functions; the first-class property alone does not make the original
constraint fields commute or provide a global closed dual coframe.
Regular momentum levels for Hamiltonian vector-group actions give
an Abelian realization of this construction.

In the $q$-cosymplectic setting, we characterized canonical fields
and related generalized canonical transformations to the
leafwise symplectic geometry and locally Hamiltonian fields.
The Noether theorem identifies quantities conserved both by
$X_H$ and by every Reeb field, hence along every direction
$X_H+\sum_i a_iR_i$.
This conservation statement does not require the directional
evolution fields themselves to commute.

In the uniform $q$-contact setting, we described canonical fields
in terms of Reeb-invariant Hamiltonians and constant Reeb
components. We also related dynamical symmetries to dissipated
$q$-tuples. For strict infinitesimal symmetries, the differences
$\lambda_i(V)-\lambda_j(V)$ are locally constant, so a single
such symmetry does not yield $q$ functionally independent
conservation laws. More general dynamical symmetries can produce
dissipated tuples with nonconstant conserved differences.
Common constant rescalings of the defining one-forms generate
canonoid transformations for every Hamiltonian; for $q\geq2$,
a common infinitesimal conformal factor is necessarily locally
constant.

For a one-parameter group of canonoid transformations in either
setting, the infinitesimal relation is
$[X,X_H]=X_{L_XH-\kappa}$, with the structural conditions on
$\kappa$ appropriate to each geometry. The counterexamples in
Sections~\ref{sec:cosymp} and~\ref{sec:contact} show that these
infinitesimal conditions do not, by themselves, ensure that the
flow consists of canonoid transformations. We also gave
sufficient conditions under which the finite-flow property holds.

The physical examples clarify the different roles of the
characteristic flows. For free relativistic particles, the chosen
clocks are proportional to the individual proper times, and the
Lorentz momentum functions are invariant under each corresponding
Reeb flow. For interacting constraint models, the existence and
physical interpretation of suitable clocks require additional
assumptions. In the monopole systems, the Reeb fields generate
gauge transformations, while the particle has a single physical
time. The rotational symmetry of the one-monopole model recovers
the Poincar\'e angular momentum.

The two-monopole model separates two global issues.
Irrationally related charges obstruct the intermediate quotient
by the characteristic flow of the total Cartan form, whose
orbits are dense in the gauge tori; the quotient by the full
gauge torus remains smooth. Independently of this irrationality,
the model carries a non-uniform $2$-contact structure, and the
topology of the bundle obstructs a uniform structure with the
prescribed gauge generators as Reeb fields.
This obstruction concerns the fixed Reeb frame, rather than
all possible uniform structures on the underlying manifold.

These results suggest the following directions.
\begin{enumerate}
\item
\emph{Non-Abelian characteristic symmetries.}
The Lie-group formulation of Hamilton--Jacobi theory in
\cite{Mukunda1978,MarmoMorandiMukunda1990} suggests retaining a
prescribed frame $V_1,\dots,V_q$ with
$[V_a,V_b]=\sum_c c_{ab}^{\,c}V_c$.
A dual family necessarily satisfies
$d\lambda_c(V_a,V_b)=-c_{ab}^{\,c}$.
Thus non-Abelian extensions require modifying the structural
conditions themselves: allowing distinct $d\lambda_i$ within
the present definition of a $q$-contact structure still gives
commuting Reeb fields.
A suitable extension should specify the compatible two-forms,
the relevant group actions and momentum levels, and the
corresponding notions of canonical and canonoid transformation.

\item
\emph{Non-uniform structures and their dynamics.}
Develop transformation theory for non-uniform $q$-contact
structures and for $q$-structured manifolds with distinct
two-forms, in conjunction with the Noether framework of
\cite{deLeon2026geometric}.
The two-monopole example requires keeping track of the physical
presymplectic form as well as the individual $d\lambda_i$.
A central question is which transformations preserve the
compatibility between these forms and the chosen dynamics.

\item
\emph{From infinitesimal conditions to canonoid flows.}
Identify geometric hypotheses that make the infinitesimal
canonoid conditions sufficient along the entire flow, and
describe the higher-order obstructions exhibited by our
counterexamples. The distinction between local Hamiltonian
functions and globally defined transformed Hamiltonians
should also be included in this analysis.

\item
\emph{Global clocks and relativistic covariance.}
Determine when interacting constraint systems admit global
closed clock forms and primitives with the normalization
required by Proposition~\ref{prop:cosymp-to-contact}.
It is also necessary to understand the action of the full
Poincar\'e group on these choices. In the free model,
translations do not preserve the particular clock forms
used here; their action between different choices of
geometric data deserves a separate treatment.

\item
\emph{Reduction, integrability and dissipative applications.}
Determine when canonical and canonoid transformations descend
through reduction, and how the additional one-forms and
Hamiltonians descend with them.
Relations with integrability should include the compatibility
conditions needed for commuting evolution fields.
Thermodynamical models such as those in
\cite{deLeonBajo2025} provide a further setting in which to
investigate these questions and the interpretation of
dissipated tuples.
\end{enumerate}

In a forthcoming paper, we shall investigate more closely
the relation between the present geometric approach
and Schwinger's quantum action principle.

\section*{Acknowledgment}
 The research of X.~Zhao is supported by NSFC (Grant No.~12401234).

\medskip
\noindent\textbf{Conflict of interest statement.} On behalf of all authors, the corresponding author states that there is no conflict of interest.

\medskip
\noindent\textbf{Data availability.} Data sharing is not applicable to this article as no new data were created or analyzed in this study.


\begin{thebibliography}{00}
\bibitem{AlekseevskyGrabowskiMarmoMichor1994}
D. Alekseevsky, J. Grabowski, G. Marmo and P. W. Michor,
Poisson structures on the cotangent bundle of a Lie group
or a principle bundle and their reductions,
\emph{J. Math. Phys.}
\textbf{35} (1994), no.~9, 4909--4927.
\url{https://doi.org/10.1063/1.530822}.

\bibitem{Almeida}
U. Almeida,
\emph{Contact Anosov actions with smooth invariant bundles},
PhD thesis, Universidade de S\~ao Paulo (2018).
\url{https://doi.org/10.11606/T.55.2018.tde-01112018-110622}.

\bibitem{AndersonBergmann1951}
J. L. Anderson and P. G. Bergmann,
Constraints in covariant field theories,
\emph{Phys. Rev.}
\textbf{83} (1951), 1018--1025.
\url{https://doi.org/10.1103/PhysRev.83.1018}.

\bibitem{AzuajeEscobar}
R. Azuaje and A. M. Escobar-Ruiz,
Canonical and canonoid transformations for Hamiltonian systems
on (co)symplectic and (co)contact manifolds,
\emph{J. Math. Phys.}
\textbf{64} (2023), 033501.
\url{https://doi.org/10.1063/5.0135045}.

\bibitem{Azuaje2}
R. Azuaje and A. Bravetti,
Scaling symmetries and canonoid transformations in Hamiltonian systems,
\emph{Int. J. Geom. Methods Mod. Phys.}
\textbf{21} (2024), 2450077.
\url{https://doi.org/10.1142/S0219887824500774}.

\bibitem{Azuaje3}
R. Azuaje and X. F. Zhao,
Canonical and canonoid transformations for Hamiltonian systems
on locally conformal symplectic manifolds,
\emph{J. Geom. Phys.}
\textbf{222} (2026), 105761.
\url{https://doi.org/10.1016/j.geomphys.2026.105761}.

\bibitem{BMSS1980}
A. P. Balachandran, G. Marmo, B. S. Skagerstam and A. Stern,
Magnetic monopoles with no strings,
\emph{Nucl. Phys. B}
\textbf{162} (1980), 385--396.
\url{https://doi.org/10.1016/0550-3213(80)90346-6}.

\bibitem{BMMNSSZ1984}
A. P. Balachandran, G. Marmo, N. Mukunda, J. S. Nilsson,
A. Simoni, E. C. G. Sudarshan and F. Zaccaria,
Unified geometrical approach to relativistic particle dynamics,
\emph{J. Math. Phys.}
\textbf{25} (1984), 167--176.
\url{https://doi.org/10.1063/1.525987}.

\bibitem{Bogdanovic}
R. Bogdanovi\'c and M. S. Gopinathan,
A canonical transformation of the Hamiltonians quadratic
in coordinate and momentum operators,
\emph{J. Phys. A}
\textbf{12} (1979), 1457--1468.
\url{https://doi.org/10.1088/0305-4470/12/9/013}.

\bibitem{Carinena}
J. F. Cari\~nena, J. Fern\'andez-N\'u\~nez and E. Mart\'inez,
A geometric approach to Noether's second theorem
in time-dependent Lagrangian mechanics,
\emph{Lett. Math. Phys.}
\textbf{23} (1991), 51--63.
\url{https://doi.org/10.1007/BF01811294}.

\bibitem{Carinena2}
J. F. Cari\~nena, F. Falceto and M. F. Ra\~nada,
Canonoid transformations and master symmetries,
\emph{J. Geom. Mech.}
\textbf{5} (2013), 151--166.
\url{https://doi.org/10.3934/jgm.2013.5.151}.

\bibitem{CarinenaRanada1988}
J. F. Cari\~nena and M. F. Ra\~nada,
Canonoid transformations from a geometric perspective,
\emph{J. Math. Phys.}
\textbf{29} (1988), 2181--2186.
\url{https://doi.org/10.1063/1.528146}.

\bibitem{CJS1963}
D. G. Currie, T. F. Jordan and E. C. G. Sudarshan,
Relativistic invariance and Hamiltonian theories of interacting particles,
\emph{Rev. Mod. Phys.}
\textbf{35} (1963), 350--375.
\url{https://doi.org/10.1103/RevModPhys.35.350}.

\bibitem{Currie}
D. G. Currie and E. J. Saletan,
Canonical transformations and quadratic Hamiltonians,
\emph{Nuovo Cimento B}
\textbf{9} (1972), 143--153.
\url{https://doi.org/10.1007/BF02735514}.

\bibitem{Dirac1950}
P. A. M. Dirac,
Generalized Hamiltonian dynamics,
\emph{Canad. J. Math.}
\textbf{2} (1950), 129--148.
\url{https://doi.org/10.4153/CJM-1950-012-1}.

\bibitem{DubrovinGiordanoMarmoSimoni1993}
B. A. Dubrovin, M. Giordano, G. Marmo and A. Simoni,
Poisson brackets on presymplectic manifolds,
\emph{Int. J. Mod. Phys. A}
\textbf{8} (1993), no.~21, 3747--3771.
\url{https://doi.org/10.1142/S0217751X93001521}.

\bibitem{deLeonBajo2025}
M. de Le\'on and J. Bajo,
A geometric description of some thermodynamical systems,
\emph{J. Phys. A: Math. Theor.}
\textbf{58} (2025), 175203.
\url{https://doi.org/10.1088/1751-8121/adcd14}.

\bibitem{deLeon2026geometric}
M. de Le\'on, C. Sard\'on and X. Zhao,
A geometric approach to the (generalized) Noether theorem
for Hamiltonian systems on (non)-uniform $q$-contact manifolds,
arXiv:2608.05681 [math.SG] (2026).
\url{https://arxiv.org/abs/2608.05681}.

\bibitem{Dirac1931}
P. A. M. Dirac,
Quantised singularities in the electromagnetic field,
\emph{Proc. R. Soc. Lond. A}
\textbf{133} (1931), 60--72.
\url{https://doi.org/10.1098/rspa.1931.0130}.

\bibitem{Dirac1964}
P. A. M. Dirac,
\emph{Lectures on Quantum Mechanics},
Belfer Graduate School of Science, Yeshiva University, New York (1964).

\bibitem{Finamore2024}
D. Finamore,
Contact foliations and generalised Weinstein conjectures,
\emph{Ann. Global Anal. Geom.}
\textbf{65} (2024), 27.
\url{https://doi.org/10.1007/s10455-024-09957-w}.

\bibitem{Finamore2024b}
D. Finamore,
Quasiconformal contact foliations,
\emph{Math. Ann.}
\textbf{389} (2024), 1575--1598.
\url{https://doi.org/10.1007/s00208-023-02687-7}.

\bibitem{LeokSardonZhao}
M. Leok, C. Sard\'on and X. F. Zhao,
Integration on $q$-cosymplectic manifolds,
\emph{J. Nonlinear Sci.}
\textbf{36} (2026), 56.
\url{https://doi.org/10.1007/s00332-026-10279-3}.

\bibitem{Leok2}
M. Leok, C. Sard\'on and X. F. Zhao,
$q$-Cosymplectic geometry, integrability and reduction,
arXiv:2509.05998 [math-ph] (2025).
\url{https://arxiv.org/abs/2509.05998}.

\bibitem{Leok2026}
M. Leok, C. Sard\'on and X. F. Zhao,
Noether-type theorems and the generalized Herglotz principle
in $q$-contact geometry,
\emph{J. Phys. A: Math. Theor.}
\textbf{59} (2026), 375203.
\url{https://doi.org/10.1088/1751-8121/aea463}.

\bibitem{MarmoMukundaSamuel1983}
G. Marmo, N. Mukunda and J. Samuel,
Dynamics and symmetry for constrained systems:
a geometrical analysis,
\emph{Riv. Nuovo Cimento}
\textbf{6} (1983), no.~2, 1--62.
\url{https://doi.org/10.1007/BF02740487}.

\bibitem{MarmoMorandiMukunda1990}
G. Marmo, G. Morandi and N. Mukunda,
A geometrical approach to the Hamilton-Jacobi form of dynamics
and its generalizations,
\emph{Riv. Nuovo Cimento}
\textbf{13} (1990), no.~8, 1--74.
\url{https://doi.org/10.1007/BF02832785}.

\bibitem{Mukunda1976}
N. Mukunda,
Symmetries and constraints in generalized Hamiltonian dynamics,
\emph{Ann. Physics}
\textbf{99} (1976), 408--433.
\url{https://doi.org/10.1016/0003-4916(76)90101-9}.

\bibitem{Mukunda1978}
N. Mukunda,
Phase space methods and the Hamilton-Jacobi form of dynamics,
\emph{Proc. Indian Acad. Sci. Sect. A}
\textbf{87} (1978), no.~5, 85--105.
\url{https://repository.ias.ac.in/25374/1/426.pdf}.

\bibitem{Negri}
L. J. Negri, L. C. Oliveira and J. M. Teixeira,
Canonoid transformations and constants of motion,
\emph{J. Math. Phys.}
\textbf{28} (1987), 2369--2372.
\url{https://doi.org/10.1063/1.527772}.

\bibitem{PerezAlvarez2024}
J. P\'erez \'Alvarez,
Symmetries and dissipation laws on contact systems,
\emph{Mediterr. J. Math.}
\textbf{21} (2024), 151.
\url{https://doi.org/10.1007/s00009-024-02695-0}.

\bibitem{Rastelli}
G. Rastelli and M. Santoprete,
Canonoid and Poissonoid transformations, symmetries and bihamiltonian structures,
\emph{J. Geom. Mech.}
\textbf{7} (2015), 483--515.
\url{https://doi.org/10.3934/jgm.2015.7.483}

\bibitem{Schwinger1970}
J. Schwinger,
Quantum theory of gauge fields,
in \emph{Elementary Particle and High Energy Physics},
eds. M. L\'evy and Ph. Meyer
(Gordon and Breach, New York, 1965), p.~189.

\bibitem{YanoIshihara1973}
K. Yano and S. Ishihara,
\emph{Tangent and Cotangent Bundles: Differential Geometry},
Monographs and Textbooks in Pure and Applied Mathematics,
Vol.~16 (Marcel Dekker, New York, 1973).

\bibitem{ZSNMMB1983}
F. Zaccaria, E. C. G. Sudarshan, J. S. Nilsson, N. Mukunda,
G. Marmo and A. P. Balachandran,
Universal unfolding of Hamiltonian systems:
from symplectic structure to fiber bundles,
\emph{Phys. Rev. D}
\textbf{27} (1983), 2327--2340.
\url{https://doi.org/10.1103/PhysRevD.27.2327}.

\bibitem{Zhao2025}
X. F. Zhao,
Canonoid transformation and master symmetries of Hamiltonian systems
on locally conformal symplectic manifold,
\emph{J. Math. Phys.}
\textbf{66} (2025), 082704.
\url{https://doi.org/10.1063/5.0255345}.

\bibitem{ZhaoLi}
X. F. Zhao and Y. Li,
Complete integrability in $q$-contact Hamiltonian systems,
\emph{Bull. Sci. Math.}
\textbf{213} (2026), 103928.
\url{https://doi.org/10.1016/j.bulsci.2026.103928}.

\end{thebibliography}
\end{document}